\documentclass{aa}  

\usepackage{graphicx}
\usepackage{txfonts}
\newcommand\isotope[2]{\textsuperscript{#2}#1}
\begin{document}

   \title{Standard solar models: a perspective from updated solar neutrino fluxes and the gravity-mode 
period spacing}

   \subtitle{}

   \author{S.J.A.J. Salmon
          \inst{1}
          \and
          G. Buldgen \inst{1}
                    \and
          A. Noels \inst{2}
                    \and
          P. Eggenberger \inst{1}
                              \and
          R. Scuflaire \inst{2}
          \and
          G. Meynet \inst{1}
          }

   \institute{ Observatoire de Gen\`eve, Universit\'e de Gen\`eve, Ch. Pegasi 51b, 1290 Sauverny, Switzerland \\
              \email{sebastien.salmon@unige.ch}
         \and
              STAR Institute, Universit\'e de Li\`ege, All\'ee du 6 Ao\^ut 19C, 4000 Li\`ege, Belgium }
   \date{Received ;  }

% \abstract{}{}{}{}{} 
% 5 {} token are mandatory
 \titlerunning{Standard solar models}
  \abstract
   {Thanks to the large and exquisite set of observations we have for the Sun, our star is by far a privileged target 
for testing 
stellar models with unique precision. A recent concern appeared with the progress in the solar surface abundances 
derivation that has led to a decrease of the solar metallicity. While the ancient 
high-metallicity models were in fair agreement 
with other observational indicators from helioseismology and solar neutrino fluxes, it is no longer the case for 
low-metallicity models. This issue is 
known as the solar problem. Recent collection of data are however promising to shed a new light 
on it. For instance, the Borexino collaboration released in 2020 the first-ever complete estimate of neutrinos emitted 
in the CNO cycle. It has reaffirmed the role of the neutrino constraints in the solar modelling process and their 
potential to explore the associated issues. In parallel, newly claimed detection of solar gravity modes of 
oscillations offers another opportunity of probing the stratification in the Sun's central layers.}
   { We propose to combine the diagnoses from neutrinos and helioseismology, both from pressure and gravity modes, for 
assessing the predictions of solar models. We compare in detail the different 
physical prescriptions currently at disposal for stellar model computations.} 
   {We build a series of solar standard models by varying the different physical 
ingredients affecting directly the core structure; opacity, chemical mixture, nuclear reactions rates. 
We compare the predictions of these models to their observational counterpart for the neutrinos fluxes, gravity-mode 
period spacing and low-degree pressure mode frequency ratios.}
   {The CNO neutrino flux confirms previous findings, showing a preference for high-metallicity models. 
Nevertheless, we found that mild modification of the nuclear screening factors can 
re-match low-metallicity model predictions to observed fluxes, although it does not restore the agreement with the
helioseismic frequency ratios. Neither the high-metallicity or low-metallicity models are able to reproduce the 
gravity-mode period spacing. The disagreement is huge, more than 100$\sigma$ to the observed value. Reversely, the 
family of standard models narrows the expected range of the Sun's period spacing: between $\sim$2150 to 
$\sim$2190~s. Moreover, we show this indicator can constrain the chemical mixture, 
opacity, and to a lower extent nuclear reactions in solar  models.}
   {}


   \maketitle

%\linenumbers

\section{Introduction}

The Sun is under intense scrutiny as a testbed of stellar physics. We benefit from a unique view 
of its internal structure thanks to solar oscillation -helioseismic- observations and solar neutrino detections, while 
we can 
estimate its envelope composition from spectroscopic determinations of element
abundances at its surface.

Helioseismology has precisely constrained the Sun's convective envelope and neighbour 
superficial 
radiative regions \citep[see e.g. recent reviews by][and ref. therein]{buldgen19,jcd20}.  In a non-exhaustive list, we 
can highlight; the 
determination of the location of the base of the 
convective 
envelope \citep{kosovichev91,jcd91}, the reconstruction of the internal rotation profile and the highlighting of the 
tachocline \citep[e.g.][]{kosovichev88,brown89,schou98}, the determination 
of the helium abundance in the convective envelope \citep{vorontsov91,basu95}, and seismic inversions of the 
sound speed profile \citep{jcd85}, as well as other structural variables \citep[see reviews by][and references 
therein]{jcd02,kosovichev11,basu16}.

In parallel, the constant improvement of detectors of neutrinos from extra-terrestrial sources, 
in particular intended to detect those of solar origin \citep[started decades ago, see][]{davis68}, has set a new path 
to constrain the 
physical conditions and nuclear burning in 
the Sun's central layers \citep[e.g, including reviews,][]{bahcall88,tc11,haxton13}. The 
potential of
solar neutrino measurements has been confirmed by supporting the evidence for neutrino 
oscillation \citep[e.g.][]{bahcall98,fukuda99}, and binding solar central temperatures, opacities, abundances or 
nuclear reaction rates \citep[e.g.][]{tc93,inno98,watanabe01,antia02,gonzalez06,serenelli13,serenelli16}. 
 Interestingly, these central temperatures can be 
compared to those estimated from seismic models \citep[e.g.][]{antia95,ricci97,antia98}, and reveal potential flaws in 
the standard solar models.

%\bfseries Helioseismic and neutrino constraints are mutually beneficial to the study of the Sun. For instance, the 
%central temperatures can be estimated with help of seismic models, which are static models built to reproduce the 
%acoustic structural quantities derived from helioseismology \citep[as e.g.][based on sound speed or density 
%inversions]{antia98}. Comparison of this ``helioseismic'' temperatures to that required to reproduce neutrino data can 
%then reveal potential flaws in the stellar models \citep[e.g.][]{antia95,ricci97}. 
% It also allow to the central temperatures in the Sun or verify the radiative luminosity and that from neutrino fluxes. 
% decades ago, the solar neurino flux problem illustrated the powerfulness of combining helioseismology and measures of 
% solar neutrino fluxes; the core temperatures and nuclear reaction rates then predicted by solar models constrained by 
% helioseismology predicted neutrino fluxes larger than those observed. Solving this issue finally give support for the 
% evidence of a neutrino flavour oscillation \citep{bahcall98,dziembowski90} .

Despite these successes, solar physics now faces a stalemate. The chemical element abundances composing the solar 
plasma are an obvious key ingredient for computing a model of 
the Sun. They are 
taken as the Sun's surface composition, derived by spectroscopic analysis of its 
photosphere. In 2005, new 3D atmosphere simulations and better atomic data led to a downward revision of the most 
abundant elements, C, N, O, and Ne, by $\sim30\%$ \citep[][hereafter AGS05]{asplund05}. The solar 
metallicity decreased in revised determinations 
\citep[AGS05;][hereafter AGSS09]{caffau11,asplund09} in comparison to previous determinations 
\citep[e.g.][]{grevesse93,grevesse98}. As a consequence, 
standard solar models (hereafter SSMs) including revised abundances are no longer in agreement with helioseismology: 
the base of the convective 
envelope is too shallow, the helium surface abundance is lower than the helioseismic one, and the results of inverse 
methods for acoustic 
variables present larger differences with the Sun's acoustic structure 
\citep[][]{montalban04,tc04,bahcall05,antia05,guzik06,serenelli09}. 
Rapidly, possible solutions or expected improvements to this issue were proposed, see e.g. \citet{basu08} or 
\citet{guzik10}; opacity underestimation, accretion by young Sun, overshooting,... 

Comparison to solar neutrino fluxes similarly shows that the SSMs divide in two categories 
according to the adopted chemical mixture; the high-metallicity ones (old solar abundance determinations) are 
favoured as they better predict the rates of production of solar neutrinos than those of low 
metallicity (revised solar abundances), see e.g. works by \citet{bergstrom16} and 
\citet{vinyoles17}. This result relies in particular on the analysis of $\Phi$(Be) and $\Phi$(B), the respective 
neutrinos fluxes produced by the $^7$Be electronic capture in the ppII branch, and the $\beta$ decay of $^8$B 
in the ppIII branch (subchains of the pp H-burning process).

However, we can now count on precision improvement and new observational constraints for shedding a 
new light 
on these issues. At first, the Borexino collaboration \citeyear{borexinoCNO20} has improved greatly the determination 
of neutrino fluxes from the CNO cycle, and  gave for the first time an estimate of the fraction of nuclear energy 
generated by CNO in the 
Sun. Recently, \citet{fossat17} announced the detection of solar gravity (g) modes from 
the analysis of 
16.5 year-long data series of the GOLF instrument dedicated to helioseismology \citep{gabriel95}, on board of the SOHO 
satellite. Besides constricting rotation in deeper layers of the Sun than pressure (p) modes, the period spacing of the 
g modes, a nearly constant value, is sensitive to the stratification at the centre \citep[e.g.][]{berthomieu91}. 
However, a series of works \citep{schunker18,appourchaux19,scherrer19} puts serious doubts on this 
recent detection, 
which is now more than weakened. But given the potentially reachable precision on the period spacing with the methods 
used in \citet{fossat17} and \citet{fossat18}, exploring the way this latter can constrain solar models remains of 
interest 
\citep[see 
e.g. comparison with estimates from seismic models in][]{buldgen20}.

We show how a firm detection of the period spacing in combination with the most recent 
solar neutrinos constraints would be strongly complementary to explore the central physical conditions of the 
Sun. We also use information on the 
innermost regions that can be given by solar p-modes, through their combination as frequency ratios 
\citep[see][]{roxburgh03,chaplin07}. We compare these indicators with a series of SSMs for 
which physics is varied following: chemical mixture, opacity, nuclear reaction rates, microscopic diffusion. These 
standard inputs are the most affecting factors of conditions at the Sun's core, and so are the best to explore with the 
observational dataset proposed above. As we focus on central layers of 
models, the outer envelope layers will not necessarily be in agreement with all of the helioseismic 
indicators.

We start in Section~2 by presenting the different solar observational constraints considered in this paper. We then
describe the series of standard solar models and their different input physics in Sect.~3. We check their 
consistency with g-mode spacing, neutrino 
fluxes, and frequency ratios of low-degree p modes in Sect.~4.  We discuss the accuracy of the results in Sect.~5 
and end with our conclusions in Sect.~6.

%--------------------------------------------------------------------
\section{Observational neutrino fluxes, gravity-mode period spacing, and pressure-mode frequency ratios}
\label{section-obs}
%-------------------------------------- Two column figure (place early!)
%Helioseismology and neutrino constraints bear different, but complementary, information helping to disclose solar 
%structure and physic. The stellar oscillations are directly depending on the acoustic properties   delivers  We give 
%in more detail below the quantities that are by these obervational constraints, and the 

The measurement of neutrinos produced by the Sun carries information on the thermal structure at its 
centre. There, the neutrino  production is a function of the nuclear reaction rates, the chemical abundances (and 
plasma density), and temperatures. Considering the neutrino 
fluxes predicted by a SSM, they not only depend on the choice of the nuclear reaction rates and element 
abundances, but also on the parameters affecting the thermal structure. This model structure depends itself on the 
nuclear reaction 
rates and abundances, but also the opacity and to a lower extent, the equation of state. In this way, it lists the 
essential physical ingredients of solar models that neutrinos afford to test.

The solar neutrino fluxes $\Phi$ that can be determined using terrestrial experiments are related to the following 
nuclear reactions and electronic captures/disintegrations, parts either of the pp chain or the CNO cycle:

%    \begin{eqnarray}
%       \frac{\pi^2}{8} \frac{1}{\tau_{\mathrm{ff}}^2}
%                 ( 3 \Gamma_1 - 4 )
%          & > & 0 \label{ZSDynSta} \\

\begin{eqnarray}
\label{eq-reactions}
&\Phi \textrm{(pp)}:& \textrm{\isotope{H}{1}} + \textrm{\isotope{H}{1}}  \rightarrow \textrm{\isotope{H}{2}} + 
\textrm{ e}^+ + \nu_{\textrm{e}} \label{eqn1} \\
&\Phi \textrm{(Be)}:& \textrm{\isotope{Be}{7}} + \textrm{ e}^-  \rightarrow \textrm{\isotope{Li}{7}} + 
\nu_{\textrm{e}} \label{eqn2} \\
&\Phi \textrm{(B)}:& \textrm{\isotope{B}{8}} \rightarrow \textrm{\isotope{Be}{8}}^* + \textrm{ e}^+ + 
\nu_{\textrm{e}} \label{eqn3} \\
&\Phi \textrm{(N)}:& \textrm{\isotope{N}{13}} \rightarrow \textrm{\isotope{C}{13}} + \textrm{ e}^+ + 
\nu_{\textrm{e}} \label{eqn4} \\
&\Phi \textrm{(O)}:& \textrm{\isotope{O}{15}} \rightarrow \textrm{\isotope{N}{15}} + \textrm{ e}^+ + 
\nu_{\textrm{e}} \label{eqn5} \\
&\Phi \textrm{(F)}:& \textrm{\isotope{F}{17}} \rightarrow \textrm{\isotope{O}{17}} + \textrm{ e}^+ + 
\nu_{\textrm{e}} \label{eqn6} \\
&\Phi \textrm{(pep)}:&  \textrm{p} + \textrm{e}^- + \textrm{p} \rightarrow 
\textrm{\isotope{H}{2}} + \nu_{\textrm{e}} \label{eqn7} \\
&\Phi \textrm{(hep)}:& \textrm{\isotope{He}{3}} + \textrm{p} \rightarrow \textrm{\isotope{He}{4}}  + 
\textrm{e}^+ + \nu_{\textrm{e}} \label{eqn8}
\end{eqnarray}

For the isotopes implied in the reactions constituent of the CNO cycle (Eqs.~~\ref{eqn4}-\ref{eqn6}), neutrino 
production is also possible by electronic capture. This latter is not included in the computation of our SSMs, but it 
has no impact on the prediction of neutrino production rates since electronic captures occur several order of 
magnitudes less than $\beta$ decays \citep[see e.g.][]{stonehill04}.

The observational constraints for the neutrino fluxes are summarized in Table~\ref{table-observations}. The results of 
two distinct analysis are taken into consideration. The first are those obtained by \citet[][hereafter 
B16]{bergstrom16}, in which the authors did a 
statistical analysis of a large collection\footnote{see details and references in the B16 paper} of solar neutrino 
experiments; data 
from cumulative experiment based on Cl or Ga detectors (Homestake, Gallex/GNO, SAGE), and from real-time detectors 
Super-Kamiokande (4 campaign phases), SNO (3 phases) and Borexino (2 phases). The approach by B16 is, excepting the 
evident dependence on neutrino oscillations parameters, almost independent of solar models \citep[see also details 
in][]{bahcall02}. We select the set of fluxes that they derived with the solar luminosity as a constraint, for it 
does not 
induce a dependence on solar models. The solar radiative luminosity is indeed a solar-model independent measurement, 
and 
for instance, B16 adopted that of \citet{frolic98}, $\textrm{L}_{\odot}=3.842 \times 10^{33}$~erg~s$^{-1}$. Using the 
luminosity reduces the uncertainties on fluxes, in particular on $\Phi \textrm{(Be)}$ and $\Phi \textrm{(B)}$, as 
energy produced by ppII and ppIII branches is much larger than the dominant -in term of number of reactions- ppI one. 

The second set we present in Table~\ref{table-observations} are the fluxes reported by the Borexino collaboration. 
The Borexino experiment is highly sensitive to low-energy neutrino and the collaboration did an intense 
effort to identify and reduce sources of background contamination. It led to a series of premieres; first measure of
$\Phi \textrm{(Be)}$, and a direct evidence of $\Phi \textrm{(pp)}$, including the measure of 
its spectra. Finally, it recently provided the first
direct measurement of the neutrinos produced by the CNO cycle (Eqs.~\ref{eqn4}-\ref{eqn6}). In comparison to B16 who 
could not include all of the Borexino campaigns, the 
latest results of the Borexino collaboration (\citeyear{borexinoPP18,borexinoCNO20}) rely on more data 
accumulation. They are of interest for comparison with SSMs, since in addition to refined 
 $\Phi \textrm{(pp)}$, $\Phi \textrm{(Be)}$ and $\Phi \textrm{(B)}$ values, the collaboration provides an 
absolute estimate of $\Phi \textrm{(CNO)}$. 

The Borexino results show a significant difference for $\Phi 
\textrm{(B)}$, 10\% larger than in B16 analysis. Although the two sets agree within 1$\sigma$ due to the large errors 
on the Borexino set, the change in the estimation of $\Phi 
\textrm{(B)}$ could impact the comparison with theoretical solar models. Moreover, as specified in 
B16, the derivation of $\Phi$($^8$B) is almost insensitive to the solar luminosity constraint. The comparison with 
results from other experimental facilities confirms that the Borexino measurement for $\Phi$($^8$B) gives the highest 
estimated value. For instance, it exceeds by $\sim8\%$ that of the two other recent neutrino experiments SNO 
\citep{SNO13} and Super-Kamiokande \citep{kamio16}, although they remain all in agreement to the 1$\sigma$ level.

%However, the values derived by the collaboration present larger errors than 
%in the B16 meta-analysis since the former did not exploit 
%the $\textrm{L}_{\odot}$ constraint.

We do not use the $\Phi \textrm{(pep)}$ observational determination \citep[e.g.][]{borexinopep} for comparisons with 
our SSMs. The pep reaction (Eq.~\ref{eqn7}) is an alternative branch to the p+p production of deuteron 
(Eq.~\ref{eq-reactions}). The pep reactions are not included in the nuclear network of our stellar models in reason of 
their marginal contribution to the total pp-chain energy production. If included in solar models (e.g. B16), only $\sim 
0.6\%$ of $\textrm{\isotope{H}{2}}$ appear to be created through pep channel. Moreover, the pep reaction rate 
shares the same nuclear matrix elements than that of the p+p reaction, so that the pep is expressed as a function of 
the p+p rate \citep[see][]{adelberger11}. It would not be actually an independent constraint, as it would rely on an 
estimate of the p+p reactions in our models. 

Similarly, we do not included $\Phi \textrm{(hep)}$  when testing the SSMs 
in Sect.~\ref{section-comparaison}. Although the hep proton capture (Eq.~\ref{eqn8})  generates the most energetic 
neutrinos, experimentally accessible, the probability of pp chain to go trough this reaction is very low 
($\sim10^{-5)}$) and is not included in our nuclear network. The hep 
reaction cross-section is difficult to compute, only accessible by theoretical mean. It suffers of a 
large uncertainty (large in comparison to the other reactions involved in H burning) of $\sim30\%$ 
\citep{adelberger11}, so that its interest for studying the structure of SSMs is low.

\begin{table}
\caption{Solar neutrino fluxes at 1 AU from  the combined analysis of B16, and from the Borexino collaboration 
\citeyear{borexinoPP18} and \citeyear{borexinoCNO20}. The  $\Phi$(CNO) from Borexino is equivalent to the sum of 
the 
$\Phi$(N), $\Phi$(O) and $\Phi$(F) fluxes. The g-mode period-spacing of Fo17 is given in the last row.}             
\centering  
\label{table-observations}

\begin{tabular}{l l l l}     %  columns 
\hline\hline       

            \noalign{\smallskip}

    Reference  & B16 & Borexino & Fo17 \\ 
                 \hline
            \noalign{\smallskip} 
    $\Phi$(pp) [$\times 10^{10}$ /cm$^2$ /s] & $5.97^{+0.04}_{-0.03}$ & $6.1^{+0.6}_{-0.7}$ &  \\
                \noalign{\smallskip} 
    $\Phi$(Be) [$\times 10^{9}$ /cm$^2$ /s] & $4.80^{+0.24}_{-0.22}$ & $4.99^{+0.13}_{-0.14}$ &  \\
                \noalign{\smallskip} 
    $\Phi$(B) [$\times 10^{6}$ /cm$^2$ /s] &  $5.16^{+0.13}_{-0.09}$ & $5.68^{+0.39}_{-0.41}$ & \\
                \noalign{\smallskip} 
    $\Phi$(N) [$\times 10^{8}$ /cm$^2$ /s] &  $5.03^{+8.58}_{-2.96}$ & &  \\
                \noalign{\smallskip} 
    $\Phi$(O) [$\times 10^{8}$/cm$^2$ /s] & $1.34^{+1.34}_{-0.89}$ & & \\
                \noalign{\smallskip} 
    $\Phi$(F) [$\times 10^{6}$/cm$^2$ /s] & $<8.5$ & & \\
                \noalign{\smallskip} 
    $\Phi$(CNO) [$\times 10^{8}$ /cm$^2$ /s] & & $7^{+3}_{-2}$ & \\
               \hline
               \noalign{\smallskip} 
                
    P$_0$ [s] & & & $2041 \pm 1$\\
            \noalign{\smallskip}
            \hline
         
   \end{tabular}      
   \end{table}

\subsection{Helioseismic indicators}

The information carried by oscillation modes is function of acoustic quantities (e.g. the sound speed, $c$) that 
differ according to the nature of the modes. They are not directly sensitive to the thermal structure as 
it is the case for the neutrino constraints. Combining these indicators thus provides an access to the solar 
structure and its associated physics under complementary views.  

The natural complement for probing central solar regions would be the knowledge of 
gravity modes. They generally propagate in the central regions of stars, but given the 
extended evanescent region constituted by the convective envelope, the g-modes are however expected to be of very low 
amplitude at the solar surface. They hence are a real challenge for detection \cite[see review by][]{appourchaux10}. 
Claimed discoveries of solar g-modes were made in the past \citep{g1,g2,garcia07} but were questioned or remain 
unconfirmed \citep[see detailed chronological review by][]{appourchaux13}. More recently, \citet[][hereafter 
Fo17]{fossat17} -see also \citet{fossat18}- announced another detection of this long-awaited helioseismic missing 
link. \citet{fossat17} present what would correspond to the signatures, each of a hundred of modes, of asymptotic 
g-modes of angular degree $\ell=1$ and $\ell=2$. With the rotational 
splittings that they determine, the core rotation of the Sun would be $\sim$ 3.8 higher than  
that of the envelope.

For the first time, Fo17 also provide a precise estimation of the asymptotic period spacing 
P$_0=2041 
\pm 1$~s, also given in in Table~\ref{table-observations}.  The almost constant value in period between g-modes of same 
$\ell$ and consecutive 
radial orders $n$ is well approximated at first order by the asymptotic period spacing $P_0$, provided a factor 
$1/\sqrt{\ell(\ell+1)}$. Following asymptotic developments 
\citep[][]{provost86,ellis86}, P$_0$ for the Sun can be expressed in good approximation as:
\begin{equation}
 \frac{1}{\textrm{P}_0}= \frac{1}{2 \pi^2} \int_{0}^{r_c} \frac{N}{r}dr
\end{equation}
where $r_c$ is the location of the base of the convective zone, $N$ the Brunt-V\"ais\"al\"a 
frequency, and $r$ the radius. The central layers weigh the more in the integral in reason of the 
variation in $1/r$. The P$_0$ is thus a good marker of the chemical stratification of the core region, since $N$ 
takes 
its largest values in layers with marked chemical composition gradients, and has early on appeared as a 
candidate for characterising the innermost regions of solar models \citep[e.g.][]{berthomieu91}. 

The detection by Fo17 is has been put into doubt \citep{schunker18,scherrer19,appourchaux19}. Early results from a 
series of seismic models derived by \citet{buldgen20} confirm a strong disagreement between the values 
predicted by these models and the Fo17 observational one. All this leads to the conclusion that the detection cannot be 
relied on. However, Fo17 reported its value with a high precision of $\sim$0.05\%. We hence compared that period 
spacing with SSMs to 
confirm its disagreement with solar models \citep[in the wake of preliminary results by][]{buldgen20}, whatever 
the physics used. But we also verify whether in combination with neutrino fluxes and assuming such a potential 
precision, $\textrm{P}_0$ helps discriminate SSMs with different physics, and in particular between high- 
and low-metallicity ones.

The well-confirmed solar p-modes can also be combined to define seismic indicators sensitive to 
deeper solar regions. \citet{roxburgh03} proposed to combine for solar-like stars the small and large 
frequency separations of low-degree p-modes as:
\begin{eqnarray}
\label{eq-ratio02}
r_{02}(n)&=&\frac{\delta \nu_{n,0}}{\Delta \nu_{n,1}} \\  
r_{13}(n)&=&\frac{\delta \nu_{n,1}}{\Delta \nu_{n+1,0}} \label{eq-ratio13}
\end{eqnarray}
where $\Delta \nu_{n,\ell}=\nu_{n,\ell}-\nu_{n-1,\ell}$ and 
$\delta \nu_{n,\ell}=\nu_{n,\ell}-\nu_{\textrm{n-1},\ell+2}$ are respectively the large and small frequency 
separations, with $\nu_{n,\ell}$ the frequency of the mode of order $n$ and degree $\ell$. The frequency ratios 
$r_{02}$ and $r_{13}$ have the advantage to be insensitive to surface effects affecting the p-mode oscillations. 
These indicators are in particular sensitive to sound speed in the central stellar layers \citep{Gough2003}, that is 
on the gradient of chemical composition. They are useful complement to neutrinos and g-modes to probe 
the physical conditions in layers at the vicinity of the solar core. \citet{chaplin07} explored extensively which 
physical quantities are probed with their help. They also confirmed, as in \citet{basu07r}, that high-metallicity SSMs 
were clearly better at reproducing the solar frequency ratios. We 
naturally include these indicators in this work for more testing on an extended series of SSMs, like the approach in 
\citet{buldgen19}. We took the 
low-degree frequencies from the solar BiSON set \citep{bison1,bison2} to compute the observed ratios.

\section{Physics of the standard solar models}
\label{Section_SSM}

We have calibrated a series of standard solar models, following the recipe suggested for 
picturing the present Sun in \citet{bahcall82}; computing the stellar evolution of 1~M$_{\odot}$ model to the present 
age 
of the sun (4.57~Gyr) and that reproduces the Sun's luminosity and radius, 1~L$_{\odot}$ and 1~R$_{\odot}$, 
as well as 
the present-day surface metallicity  (Z/X)$_{\textrm{s}}$ (relative to X, the hydrogen abundance). This latter quantity 
depends of course on the compilation of stellar surface abundances adopted. We took for the solar luminosity the value 
recommended in 2015 by the International Astronomical Union -IAU- in resolution B3, i.e. L$_{\odot}=3.828 \times 
10^{33}$~erg~s$^{-1}$.

All of our models were computed with help of the 
Li\`ege stellar evolution code, CLES \citep{cles}. Convection is treated under the 
mixing-length theory, implemented as in \citet{cox68}.  Excepting explicit mention, we include 
microscopic diffusion with coefficients derived from the resolution of Burgers' equations following the method in 
\citet{thoul94}. Metals heavier than He are all assimilated as Fe. No convective overshooting was considered.

Unless a change in one of the ingredient is specified, the models adopt nuclear reaction rates from 
the \citet{adelberger11} compilation, FreeEos equation of state \citep{irwin12}, OPAL opacities \citep{iglesias96} 
supplemented with Potekhin's electron-conduction opacities \citep{cassisi07}, and
grey model atmosphere with Eddington's law for the temperature $T(\tau)$ relation, with atmospheres extending up to 
an optical 
depth  $\tau=10^{-4}$. The default chemical mixture is that of AGSS09. The opacities are all 
supplemented at low-temperature conditions by those of \citet{ferguson05}, adapted to the chemical mixture selected.  
Finally, the adiabatic frequencies for the computation of frequency ratios are obtained with the Li\`ege oscillation 
code, LOSC \citep{losc}.

We describe below the physics that we vary to calibrate different SSMs. We briefly review the main differences 
between datasets of stellar physics at our disposal, as well as the 
uncertainties still affecting them.

\textbf{Solar chemical mixture.} In reason of diffusion processes \citep{jcd93b}, the present-day solar surface 
composition depends on 
its past evolution. Therefore, the determination of the initial composition of the Sun stems from solar 
calibrations, and depending on the set of surface abundances, it will result in different initial values. For 
instance, calibrations based on older determinations of the Sun's surface abundances yield high metallicity estimates 
($Z \sim 0.017-0.020$). Those based on more recent abundances give low estimates ($Z \sim 
0.013$). Among the ``old'' determinations, the most 
frequently used are those of \citet{grevesse93}, hereafter GN93, and 
\citet{grevesse98}, hereafter GS98. The consequences of revision of these abundances on the solar structure 
and its helioseismic constraints has been extensively investigated in e.g. 
\citet{bahcall05,bahcall06,basu08}.

Currently, solar observational neutrino constraints tend to favour the high-metallicity SSMs 
\citep[B16]{vinyoles17,song18}, 
while the helioseismic picture is unclear; frequency ratios are better reproduced by high-metallicity models 
\citep{basu07r}, but 
seismic inversions of metallicity points to lower estimates of the metallicity in the envelope 
\citep{vorontsov13,buldgen17b}. These latter confirm modern 
determinations of surface abundances 
do not support the older chemical mixtures \citep[e.g.][]{grevesse13}. Revise determinations benefit in 
particular of 3D (vs 1D previously) hydrodynamical simulations of the solar atmosphere, thorough review of 
oscillator strengths for the computation of spectroscopic lines, and appended lists of line blends in solar 
photospheric spectra. The determination of solar abundances by \citet{asplund09}, hereafter AGSS09, appears now as a 
stable reference. The latest updates have not significantly affected the recommended abundance values 
\citep{mix20-N,AGSS15-2,AGSS15-3}, particularly in the sensitive case of C and N elements \citep{mix19-C,mix20-N}, 
which are among the highest abundant metals and so impact the metallicity.

Meanwhile, \citet{caffau11}, hereafter Caffau11, carried out an independent determination of solar abundances. 
The authors restricted their analysis to a lower number of surface 
abundances, focusing on the most abundant elements. The solar metallicity they get is between that of GS98 and 
AGSS09.

Among the most abundant metals, the determination of Ne abundance falls apart, since it cannot be derived from 
spectroscopy of the photosphere. Determined from 
quiet regions in the solar corona, recent studies by \citet{landi15} and \citet{young18} recommend an increase of 
[Ne/O] (neon to oxygen abundance) by 40$\%$. Neon contributes significantly to the opacity in solar radiative 
regions \citep[see an illustration of its contribution to solar opacities in][]{blancard12}. It is 
hence worth combining the AGSS09 set with the recommended 
increase of the Ne abundance; we refer to this mixture as AGSS09+Ne. As a summary, we compared the role of the 
chemical mixture with help of SSMs calibrated with: GN93, GN98, Caffau11, AGSS09, and AGSS09+Ne. 
%We do not enter into  more advanced scenario where the internal composition of the Sun would have been affected by 
%event in its past, like for e.g. by accretion events. We will focus on the impact of the different chemical mixture,

\textbf{Opacity.} The two opacity libraries mostly used in stellar models are those of the OPAL \citep{iglesias96} and 
OP \citep{badnell05} projects. In the solar case, attention was drifted on differences between the opacity 
datasets at conditions corresponding to the base of the convective zone (BCZ). \citet{seaton04} show these 
differences, which rise to $\sim 5\%$ at the BCZ (log T$\sim$6.3), find their origin in the equations of state 
internal to the opacity codes (OPAL predicting more metals in excited 
states than OP). 

On the quest for solving the solar problem created by the revision of metallicity, the accuracy of 
theoretical opacity data at solar conditions was seriously questioned by \citet{bailey14}. In an experimental set-up on 
the Sandia Z-pinch machine, they reproduced conditions of ionisation and temperature of the BCZ, and 
measured a much larger iron spectral opacity than predicted by theoretical opacity computations. The source of the 
discrepancy received a lot of attention, see e.g. \citet{iglesias15}, \citet{pain15} or \citet{nahar16}. Additional 
experimental campaigns at Sandia have shown other discrepancies with theoretical spectral 
opacities for iron-group elements Cr and Ni \citep{nagayama19}. An explanation to these opacity 
issues is still pending. However, it has led to an effort for renewed improvement 
of stellar opacities. The Los Alamos group used their newly developed equation of state and own set of 
atomic computations to release the OPLIB opacity 
library, specially designed for stellar evolution codes \citep{colgan16}. Despite restricted to tighter ranges of 
density and temperature conditions, the OPAS dataset \citep{mondet15} covers enough ranges of parameters for 
computation of 
solar models \citep{lepennec15}. OP, OPAL et OPAS tables present 
differences of a few $\%$ in the radiative solar regions, likely due to the internal equations of state of 
the different codes. OPLIB stands out by much lower opacities,  
$\sim 10-15 \%$ in comparison to the other tables. It leads OPLIB to impact considerably the temperature gradient of 
the radiative region. This is expected to penalise the neutrino fluxes predicted by OPLIB SSMs \citep{song18}.
We explore the role of opacity by calibrating a series of four SSMs with all the opacity tables currently at 
our disposal; OPAL, OP, OPLIB and OPAS. 

   \begin{figure*}
\includegraphics[width=9cm]{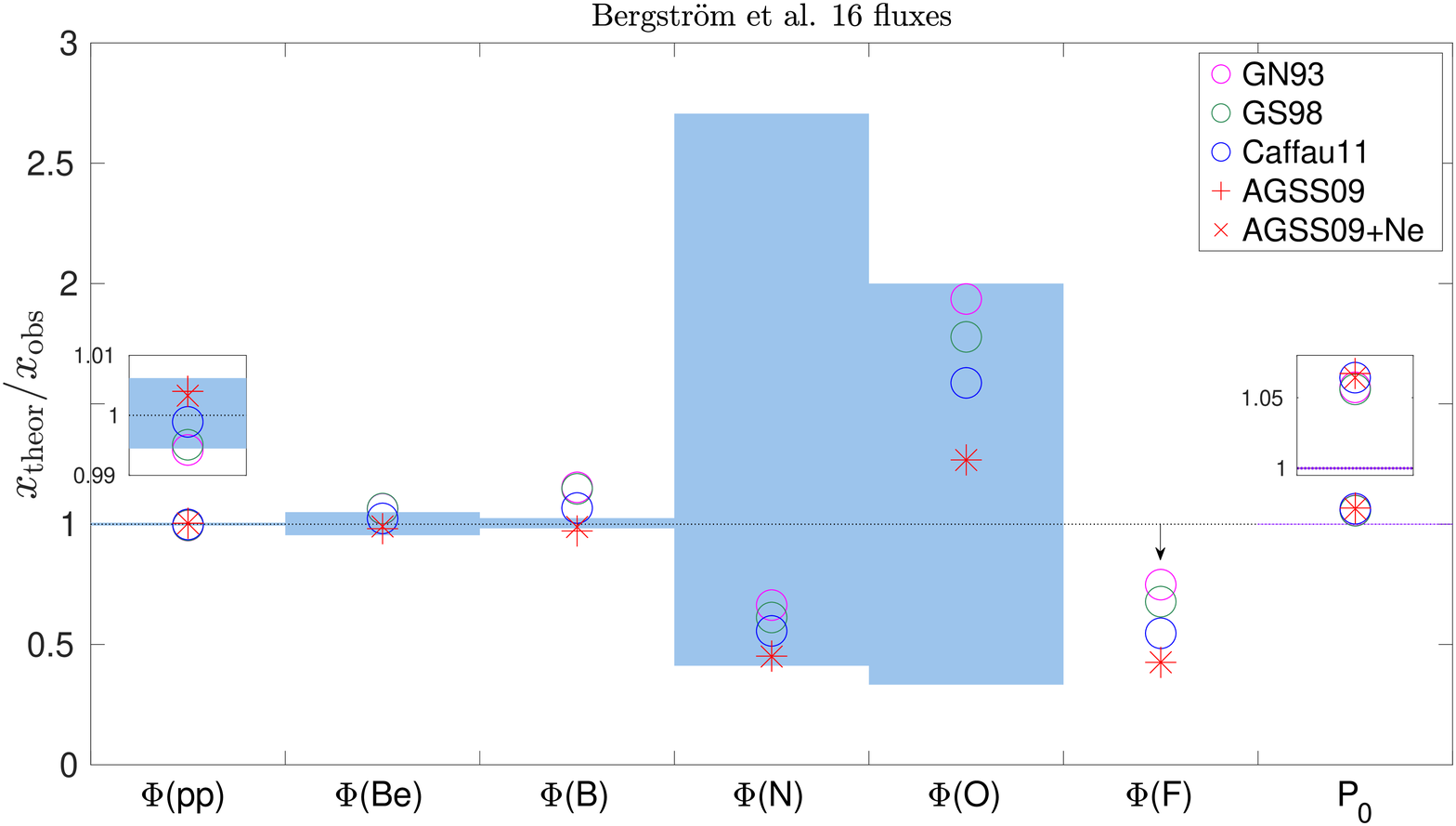}\hspace{1cm}
\includegraphics[width=9.1cm]{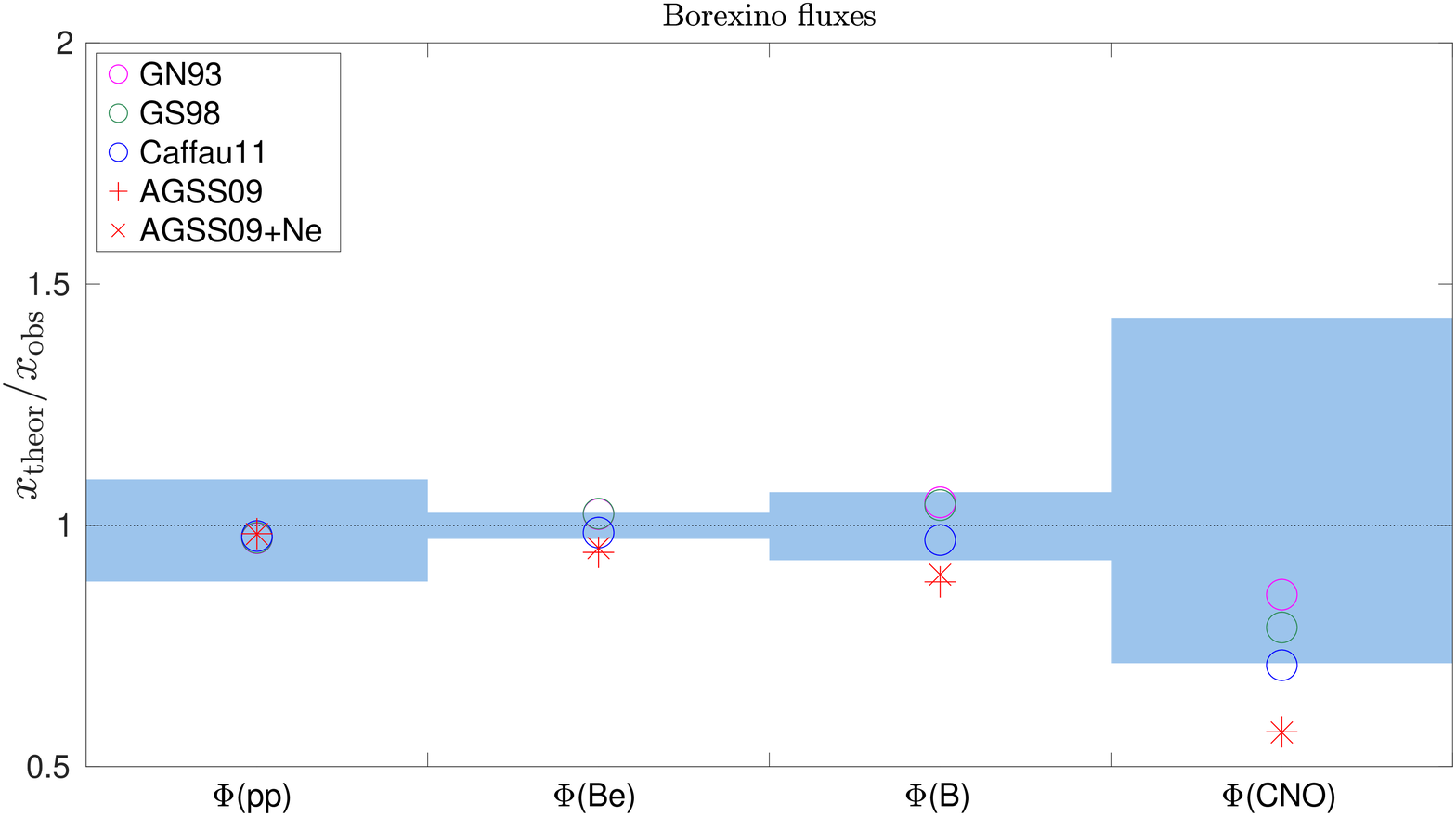}
\\
\ 
\\
\includegraphics[width=9.1cm]{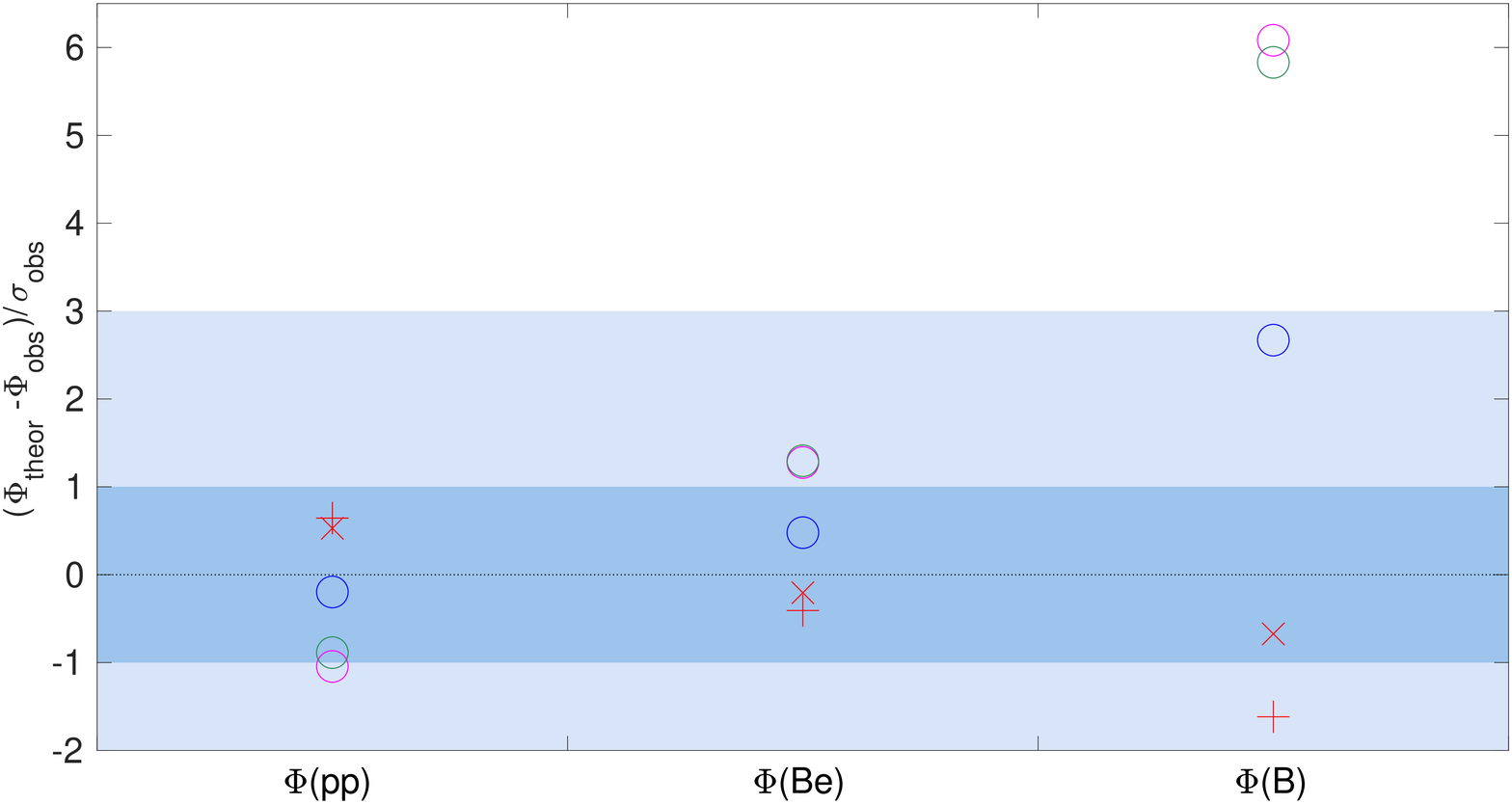}
      \caption{Neutrino fluxes predicted by the standard solar models (see Sect.~\ref{Section_SSM}) 
with different chemical mixtures. Comparison to the observational fluxes derived by B16 are shown in left panels, and 
to the Borexino 
collaboration in the right panel. In the left upper panel, two boxes are inserted to present a zoom on the $\Phi$(pp) 
and 
P$_0$ comparisons. In the panel at bottom, the comparison is restricted to $\Phi$(pp), $\Phi$(Be) and 
$\Phi$(B) from the B16 set 
only for the sake of clarity. The 1$\sigma$ intervals on the observations are shaded in 
blue in the three panels. The 3$\sigma$ range is shaded in light blue in the bottom panel.}
         \label{Fig_flux_mixtures}
   \end{figure*}

\textbf{Nuclear reaction rates.} The methods for computing astrophysical S-factors require complex nuclear 
computations; either for extrapolating results of experiment measurement (which cannot access energy domain of 
nuclear reactions in stars), with an analysis of systematics and other sources of experimental errors, or for deriving 
them completely ab initio when no experiment is feasible at all. Detailing the 
uncertainties affecting the S-factor determinations is out of the scope of this paper, and we refer to the 
thorough review of that subject by \citet{adelberger11}, also referred as the SF-II (Solar Fusion) project. Their work 
considered a whole set of nuclear reactions of 
stellar interest and is complete for those involving hydrogen burning. The S-factors of this compilation are of 
practical use for stellar computations and we selected them as the default choice for the calibration of our SSMs. We 
also calibrated two SSMs 
with help of the Nuclear Astrophysics Compilation of REaction (NACRE) rates. One calibration was made with the 
NACRE rates \citep{angulo99}, excepting the $\textrm{\isotope{N}{14}}(\textrm{p},\gamma)\textrm{\isotope{O}{15}}$ 
reaction, which follows the revision by \citet{imbriani05} (see detail below). Although it has been updated in the 
meantime, 
NACRE was used in many 
stellar models, and we thus find it interesting to confront it with solar data. We eventually considered the 
updated rates of the project, known as the NACRE II compilation \citep{xu13}. NACRE II includes the experimental 
results published in the interval of the NACRE publication. It also follows a distinct method than the results 
of R-matrix computations presented in \citet{adelberger11}. NACRE II authors extrapolate by themselves S-factors at 
low-energy following one systematic approach, based on the potential model method.

We briefly mention the estimated orders of uncertainties affecting the rates of the reactions involved in the 
production of neutrinos considered in this work (Eqs.\ref{eq-reactions}-\ref{eqn6}). These uncertainties are discussed 
in detail in 
\citet{vinyoles17}, which also include results obtained posteriorly to the reviews mentioned above. 

The p+p reaction 
(Eq.~\ref{eq-reactions}) can only be determined by ab initio computation. The details of these are nowadays well 
understood, and the error on the S$_{11}$ factor is estimated to $\sim 1\%$ \citep{adelberger11}. According 
to \citet[][and references therein]{vinyoles17}, the errors on S$_{17}$ -the reaction producing 
$\textrm{\isotope{B}{8}}$ isotopes, source of Eq.(\ref{eqn2})- is $\sim 5\%$. In the CNO cycle, details of the  
$\textrm{\isotope{N}{14}}(\textrm{p},\gamma)\textrm{\isotope{O}{15}}$ reaction are crucial. The reaction is the slowest 
and represent a bottleneck for the cycle when at equilibrium; most of the isotopes are then under the form of 
$\textrm{\isotope{N}{14}}$. Experimental measurements at the LUNA accelerator \citep[e.g.][]{formicola04,marta08} have 
recently led to an 
important reassessment of the S114 factor. Meanwhile the value recommended by SF-II, further 
measurements were conducted at LUNA \citep{marta11}. This latter suggests a decrease of $\sim 6\%$ 
of the S114 factor, although taking into account the errors, the results based on previous LUNA campaigns remain in 
good agreement. Besides, the different computational method used by NACRE II also leads to a difference of $\sim 8\%$ 
with SF-II.  

Screening effects -due to the free electron cloud reducing the Coulomb barrier between nuclides- are treated following 
the weak-screening formalism of \citet{salpeter54}. Various criticisms on the accuracy of this formalism 
have been made \citep[e.g.][]{dzitko95,shaviv01}, although to which extent it can affect screening factors is a matter 
of debate \citep{gruzinov98,bahcall02}. The 
development of an advanced formalism accounting for them is a hard task. The potential role of dynamical 
effects in the screening computation has in particular been advanced by \citet{shaviv04}. Preliminary 
attemps to include them confirm they can alter the values of the screening factor 
\citep[e.g.][]{mao09,mussack11,wood18}. Estimation of the uncertainties associated to the 
non-inclusion of dynamical effects in screening factors shows they could go up to $4-5 \%$ for some of the reactions in 
the pp chain \citep{shaviv07,shaviv10}. As screening effects play a catalyst role on nuclear reactions, we
evaluated the impact of uncertainties by implementing parametric changes of the screening factors 
in Sect.~\ref{compa-nuclear}.  

\textbf{Microscopic diffusion.} The surface metallicity along the stellar evolution of a solar model is obviously 
altered 
by diffusion. Since this metallicity is used to constrain the calibration of SSMs, the prescription of microscopic 
diffusion plays an important evolutionary effect on the resulting model. It also acts on a structural side, by 
affecting the mean molecular weight under the convective region. All our SSMs include microscopic diffusion 
based on \citet{thoul94} as mentioned above. In this approach, the perfect gas equation is assumed valid and 
the stellar plasma is considered as completely ionised. However, these hypotheses are not entirely appropriate for the  
whole solar interior, especially at low temperatures, and it can lead to overestimation of diffusion coefficients. 
To estimate this impact, we did a calibration including collision integrals in 
the diffusion
coefficients, as proposed by 
\citet{paquette86}, for accounting of departures to perfect gas conditions. In that case, oxygen was taken as the mean 
representative of metals.
%We also did another SSM calibration, accounting in the diffusion coefficient computation the 
%effect of considering partial ionization of the heavy elements (without including above collision integrals).

\textbf{Equation of state.} In first approximation, the equation of state in central radiative layers should 
be that of a perfect gas, with an adiabatic index $\Gamma_1=\partial \ln P / \partial \ln \rho |_S \simeq 5/3$, where 
$P$, $\rho$, $S$ are respectively the pressure, density and entropy. However, helioseismic inversions of this index 
revealed small departures of $\sim$0.1-0.2 \% in the deepest layers of the Sun, which are due to relativistic effects 
\citep{elliott98}. While these departures to the perfect gas are likely to have negligible impact on neutrino 
production, they affect helioseismic indicators. We tested various 
equations of state derived following the ``chemical'' picture, a method based on the minimisation of 
the free-energy. Approximations at certain levels of the computations (e.g. the effect described above) can drive 
differences between the equations of state derived following that approach. In addition to FreeEOS, used as reference, 
we did calibrations with the CEFF 
\citep[][]{eggleton73,jcd92}, and SAHA-S
\citep{saha-s-1,saha-s-2} equations of state. We also computed one SSM with the OPAL equation of state 
\citep{rogers02}, which follows the ``physical'' picture, a formalism describing the plasma elements with their
fundamental constituents and based on ab initio wavefunction computations.

\section{Comparison of model predictions with observations}
\label{section-comparaison}

We present in this section the predictions of the SSMs calibrated with the different physics aforementioned. For 
comparison 
with the two sets of observational neutrinos fluxes considered, we 
have computed reduced $\chi^2$ functions:
\begin{eqnarray}
\label{chi2}
\chi^2_{\mathrm{tot}} & = &  \chi^2_{\mathrm{neutrino}} + 
\chi^2_{\mathrm{seismo}} \\
&=& \frac{1}{\mathrm{N}_{\mathrm{tot}}-\mathrm{M}_{\mathrm{fp}} } 
\left(\sum_{i=1}^{\mathrm{N}_{\mathrm{neutrino}}} 
\frac{(\Phi_{\mathrm{obs,i}}-\Phi_{\mathrm{th,i}})^2}{\sigma_i^2} + \sum_{i=1}^{\mathrm{N}_{\mathrm{seismo}}} 
\frac{(r_{\mathrm{obs,i}}-r_{\mathrm{th,i}})^2}{\sigma_i^2} \right) \nonumber
\end{eqnarray}
where $\mathrm{N}_{\mathrm{tot}}=\mathrm{N}_{\mathrm{neutrino}}+\mathrm{N}_{\mathrm{seismo}}$, the sum of the number of 
neutrino fluxes and the number of seismic frequency ratios respectively considered. 
$\mathrm{M}_{\mathrm{fp}}$ is the number of stellar parameters let free in the calibration of the SSM. The
$\Phi_{\mathrm{obs}}$ and $\Phi_{\mathrm{theor}}$ are the solar neutrino fluxes observed and predicted by the 
theoretical SSM; $r_{\mathrm{obs}}$ and $r_{\mathrm{th}}$ are the observed and theoretical frequency ratios 
($r_{02}$ and $r_{13}$). The $\sigma_i$ are the errors associated to the corresponding observed quantities. In the case 
of the B16 data, we only included $\Phi$(pp), $\Phi$(Be), $\Phi$(B) in the 
computation of 
the $\chi^2$ function, given the large uncertainties on the fluxes from CNO. We nevertheless accounted 
for $\Phi$(CNO) in the Borexino case.

We did not take into account P$_0$ in the merit function. As we will detail in the following subsections, 
P$_0$ as determined by Fo17, actually appeared in strong disagreement with the theoretical values of our SSMs.

\subsection{Impact of the solar chemical mixture}
\label{section-mixtures}
The core temperature and central abundances, the neutrino fluxes, and the P$_0$ predicted by the SSMs 
calibrated with different mixtures are presented in Table~\ref{table-mixtures}. The immediate result is 
the confirmation of the extreme disagreement with the P$_{0}$ value reported by Fo17. As shown in the upper 
left 
panel of 
Fig.~\ref{Fig_flux_mixtures}, the values from the 
different SSMs range from 2155 to 2178~s, which is more than 100s, and so 100$\sigma$, away to that of Fo17. This 
disagreement is in support of the dispute about this detection \citep{schunker18,scherrer19,appourchaux19,buldgen20} . 
The order of the disagreement is of similar amplitude when changing other physics input in the models (see following 
subsections). The solar models cannot be used to 
interpret the nature 
itself of the signal found by Fo17, but it confirms an issue with the reported value of the 
period spacing. 

Interestingly, the range predicted by these SSMs could reversely serve as a predictive marker for refined search of 
solar g modes. Furthermore, P$_0$ varies between 10 and 20~s between the SSMs, an order larger than the observational 
precision offered by the Fo17 method. A confirmed 
detection of the solar g-mode period spacing would be an 
additional strong constraint on the central layers. The variations in P$_0$ that we observe 
in our SSMs find their origin in changes induced on $N$ in the most central layers. 
In these regions one can assume in good approximation the plasma as a perfect fully ionised gas, so that the 
Brunt-V\"ais\"al\"a frequency can be expressed as $N \simeq g^2 (\rho/P) (\nabla_{\textrm{ad}}+ \nabla \mu - \nabla 
T)$, where $g$ is the local gravity acceleration, $\nabla_{\textrm{ad}}$ the 
adiabatic gradient, $\nabla \mu$ the gradient of mean molecular weight, and $\nabla T$ the temperature gradient. 

We have consequently checked the profiles of $N$ in the five SSMs with the different adopted solar mixtures. The global 
shape of $N$ as 
a function of $r$ does not change significantly. Yet, the values of the peak in $N$ close to the centre 
($r/R\sim0.1$) do differ; among the terms in the expression of $N$ given above, we identify variation of $\nabla \mu$ 
as the largest contributor to change in $N$ and so P$_0$. The central $\nabla \mu$ increases as the 
chemical mixture is more metal-rich, leading for the GN93 and GS98 models to a lower P$_0$. The $\mu$ gradient 
appears in reason of the nuclear reactions and the fact their rates present different temperature law 
dependences. In the case of a change in the chemical mixture, the sharpness of $\nabla \mu$ is 
essentially affected by the impact of abundance modifications on the nuclear reactions. This is seen in 
Table~\ref{table-mixtures} by the variations between SSMs of X$_\textrm{c}$ and 
Z$_\textrm{c}$, the central mass fraction of H and metals. The only exception concern the difference in P$_0$ between 
the AGSS09 and AGSS09+Ne models; the change in $\nabla T$ is then dominating the changes in $N$ and P$_0$. The effect 
is not surprising as Ne is a significant contributor to opacity ($\kappa$) in the radiative layers of the 
Sun \citep[e.g.][]{antia05,lin07}.

\begin{table*}

\caption{Stellar parameters of the SSMs calibrated with different solar chemical mixtures. The neutrino 
fluxes and P$_0$ that they predict are indicated. Last rows give the reduced $\chi^2$ values 
based only on the neutrino fluxes from B16 or Borexino, as well as the total $\chi^2$, including the frequency ratio 
contributions.}            
\centering    
\label{table-mixtures}
\begin{tabular}{l l l l l l}     %  columns 
\hline\hline       

            \noalign{\smallskip}

    Solar calibration & AGSS09 & AGSS09+Ne & Caffau11 & GS98 & GN93 \\
%                 \noalign{\smallskip}
%                 & (reference calibration) & & & & \\ 
    \hline
                \noalign{\smallskip} 
     
    X$_{\textrm{c}}$ (X$_{\textrm{0}}$) & 0.356 (0.719) & 0.355 (0.717) & 0.349 (0.712) & 0.342 (0.705) & 0.342 (0.705) 
\\
                                            
                \noalign{\smallskip}        
    
    Z$_{\textrm{c}}$ (Z$_{\textrm{0}}$)  & 0.0162 (0.0151) & 0.0167 (0.0155) & 0.0185 (0.0173) & 0.0202 (0.0189) & 
0.0214 (0.0200) \\
    
    \noalign{\smallskip}

    T$_\textrm{c}$  [$\times 10^6$K] & 15.54 & 15.56 & 15.62 & 15.68 & 15.69\\
    \hline
                \noalign{\smallskip} 
         
   $\Phi$(pp) [$\times 10^{10}$ /cm$^2$ /s] &    5.995 &   5.991 &   5.965 & 5.942 & 5.937   \\
               \noalign{\smallskip}                                                                    
   $\Phi$(Be) [$\times 10^{9}$ /cm$^2$ /s]  &    4.710 &   4.755 &   4.915 & 5.111 & 5.106   \\
               \noalign{\smallskip}                                                                    
   $\Phi$(B) [$\times 10^{6}$ /cm$^2$ /s]   &    5.015 &   5.099 &   5.507 & 5.918 & 5.951   \\
               \noalign{\smallskip}                                                                    
   $\Phi$(N) [$\times 10^{8}$ /cm$^2$ /s]  &    2.273 &   2.264 &   2.801 & 3.081 & 3.340   \\
               \noalign{\smallskip}                                                                    
   $\Phi$(O) [$\times 10^{8}$/cm$^2$ /s]   &    1.695 &   1.695 &   2.123 & 2.379 & 2.590   \\
               \noalign{\smallskip}                                                                    
   $\Phi$(F) [$\times 10^{6}$/cm$^2$ /s]    &    3.619 &   3.622 &   4.648 & 5.765 & 6.365   \\
       \hline 
                   \noalign{\smallskip}  
   P$_0$ [s] & 2178 & 2172 &  2172 & 2155 & 2158   \\
    \hline 
                \noalign{\smallskip}

       $\chi^2_{\textrm{neutrino-B16}}$  &  0.033 &   0.009  & 0.142  & 0.693 & 0.754  \\

                      \noalign{\smallskip}  
              $\chi^2_{\textrm{neutrino-Borexino}}$    & 0.118  &  0.091 &  0.016   & 0.020 & 0.022 \\
          
                      \noalign{\smallskip}                           
     $\chi^2_{\textrm{tot-B16}} $        & 73.024 &  41.770 &   36.669 & 8.664 &  
7.066 
  \\

                        \noalign{\smallskip}                           
     $\chi^2_{\textrm{tot-Borexino}} $        & 71.081 &  40.692 &   35.529 & 7.771 & 6.157

     \\
  
    \noalign{\smallskip}
    \hline       

   \end{tabular}      
   \end{table*}

      \begin{figure}
\centering
\includegraphics[width=9.4cm]{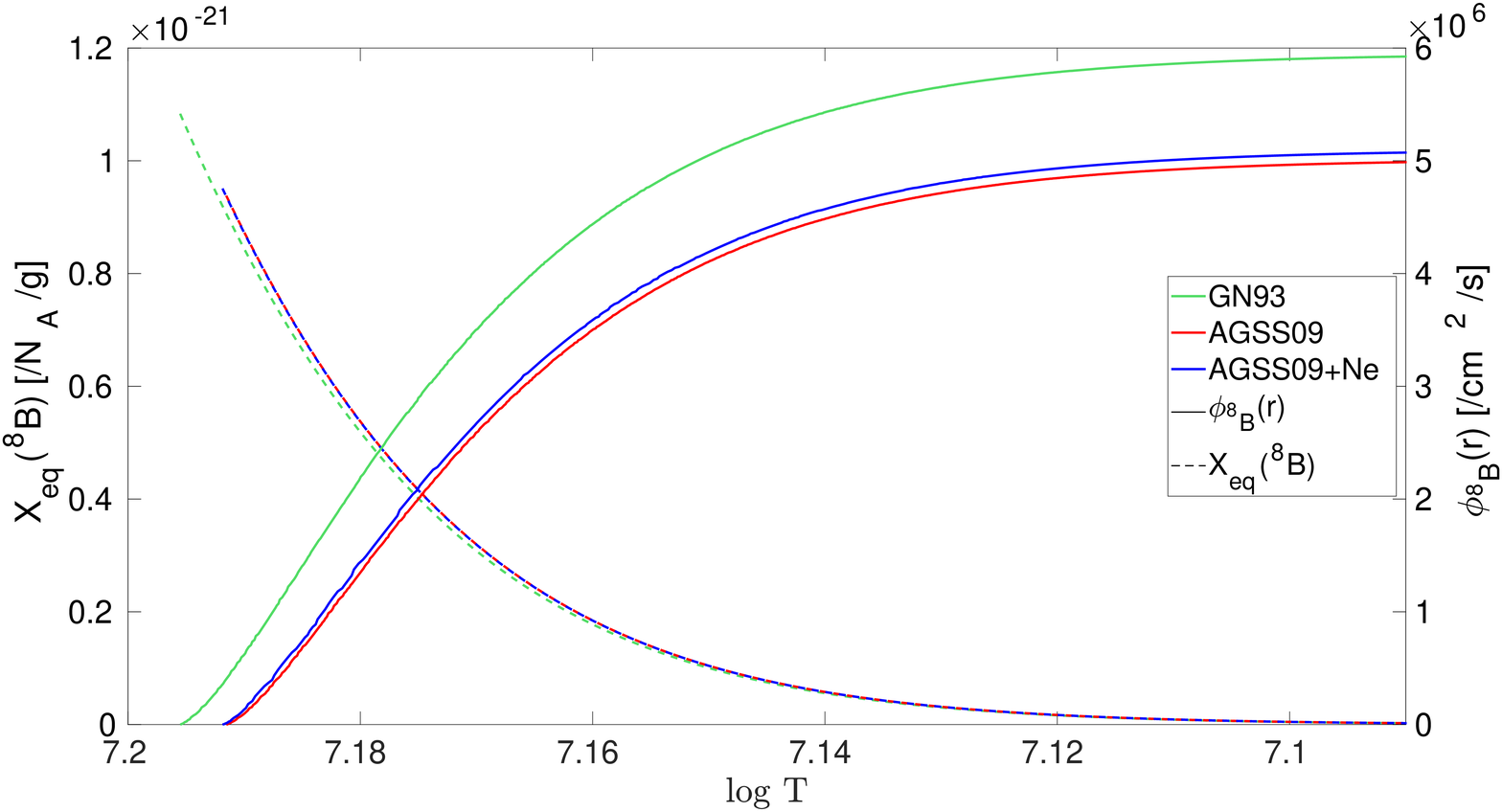}
      \caption{Equilibrium abundances (dashed lines) of the $^8$B nuclides and the cumulative neutrino flux from its 
disintegration, $\phi_{^8\textrm{B}}$ (solid lines), along the radial 
coordinate $r$. They are presented as a 
function of the temperature, for three SSMs with different chemical mixtures, as given in the legend. Abundances are in 
mole per gram, and the flux, as usual, is computed for an observer at 1AU.}
         \label{Fig_fluxB8_mixtures}
   \end{figure}   

         \begin{figure}
\centering
\includegraphics[width=9.4cm]{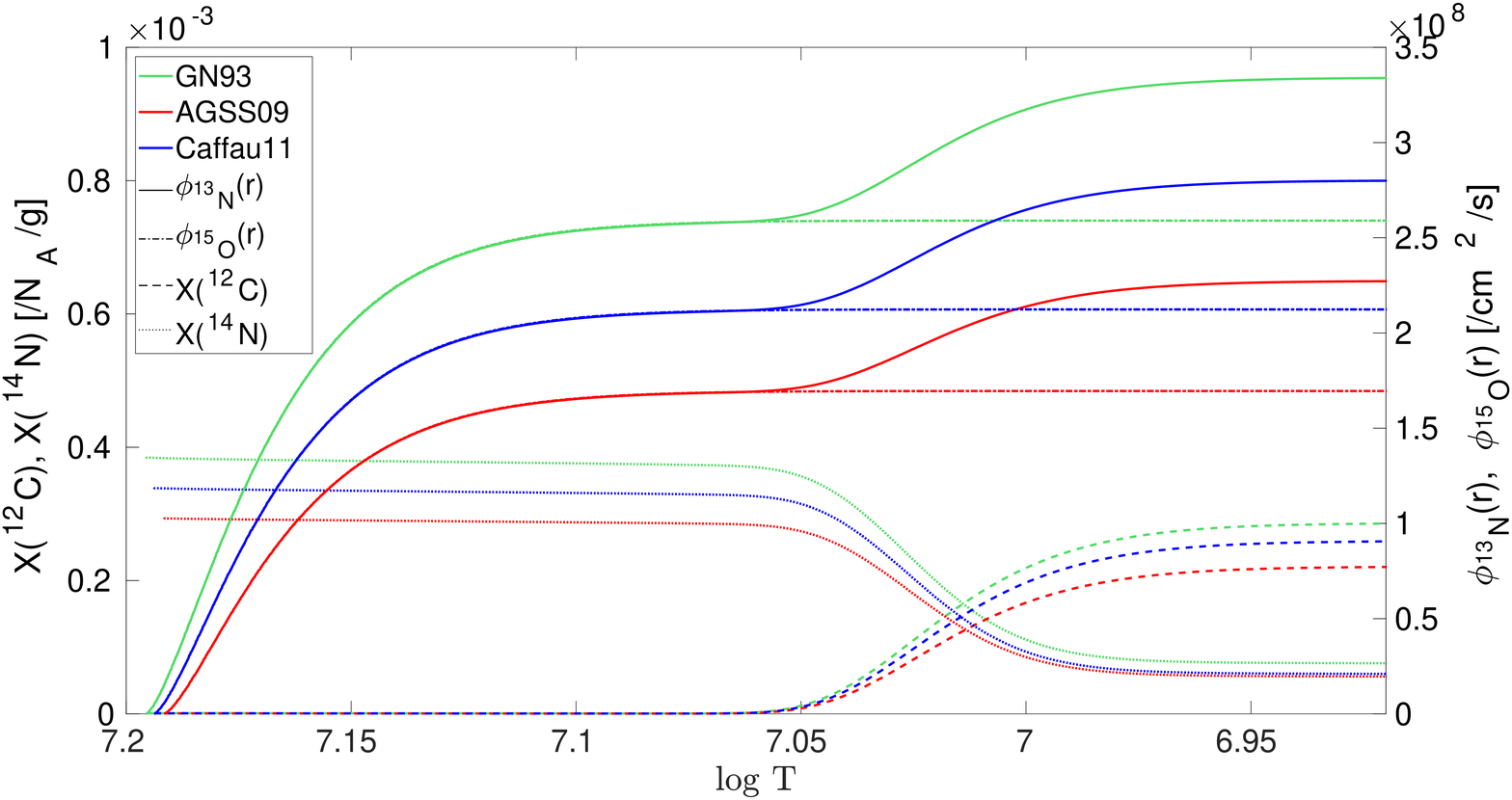}
      \caption{Equilibrium abundances of the $^{12}$C and $^{14}$N nuclides (dashed and dotted lines) 
and the cumulative neutrino fluxes of the $^{13}$N and $^{15}$O 
disintegrations, $\phi_{^{13}\textrm{N}}$ and $\phi_{^{15}\textrm{O}}$ (solid and dot-dashed lines). They are drawn as 
a function of the temperature, for 
three SSMs with different chemical 
mixture, as given in the legend.}
         \label{Fig_fluxCN_mixtures}
   \end{figure}  
   
%         \begin{figure*}
%\centering
%\includegraphics[width=18cm]{profiles-acoustic-solarmixtures}
%      \caption{Comparison of acoustic.}
%         \label{Fig_acoustic_mixtures}
%   \end{figure*}
   
   \subsubsection{Comparison to neutrino observations: the B16 set}
\label{subsection-b16-mixtures}   

Recent studies have discussed the chemical mixture impact on neutrino predictions from SSMs and compared them with the 
B16 observational set. \citet{vinyoles17} and \citet{zhang19}  found the observed values of 
$\Phi$(pp), $\Phi$(Be) and $\Phi$(B) fall between those predicted by the SSMs they calibrated with GN98 
and AGSS09\footnote{in \citet{vinyoles17}, they actually use the AGSS09 mixture but with meteoritic abundances 
preferred for 
refractory elements}. The concordance of all their SSMs with these solar fluxes is within 3$\sigma$, although the GS98 
models are in closer agreement. 

      \begin{figure*}
\centering
\includegraphics[width=14cm]{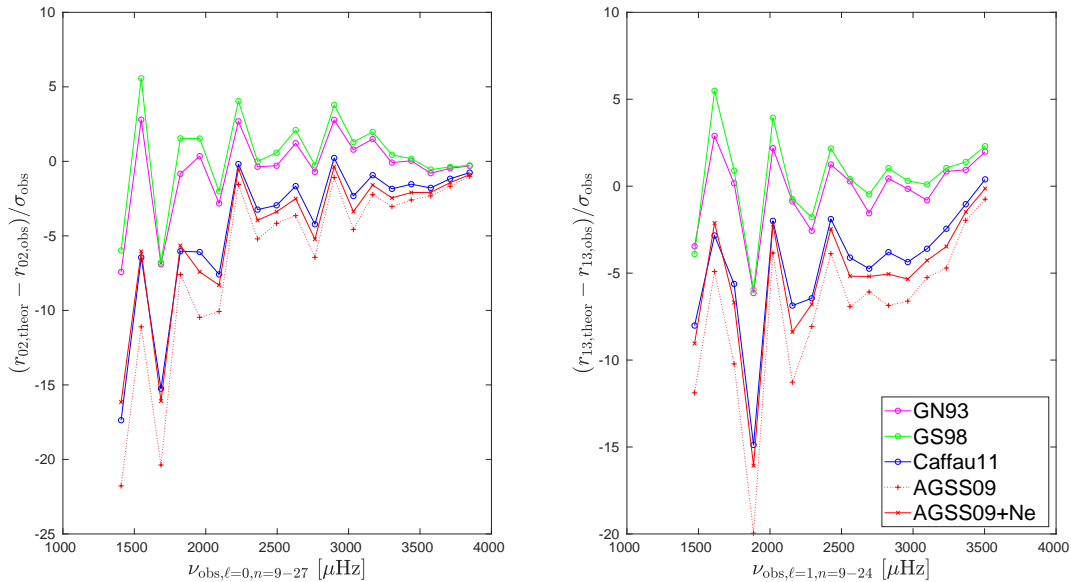}
      \caption{Comparison of solar low-degree p-mode frequency ratios r$_{02}$ and r$_{13}$ to those of the SSMs 
with different chemical mixtures. The adopted chemical mixtures are indicated in the legend.}
         \label{Fig_ratios_mixtures}
   \end{figure*}

%In the present section, the SSM models share similar physics, i.e. 
%OPAL opacities  \citet[excepting][which uses OP]{vinyoles17} and the same nuclear rates, from  
%the \citet{adelberger11} compilation. 

However, our results lead to a different picture; our SSMs with AGSS09 and AGSS09+Ne are better than the GN93 or GS98 
ones at reproducing the 
solar fluxes according to the values of $\chi^2_{\textrm{neutrino},B16}$ in Table~\ref{table-mixtures}. If we look at 
the left panels of Fig.\ref{Fig_flux_mixtures}, $\Phi$(pp) and $\Phi$(Be) are reproduced close-to or at 1$\sigma$ by 
the SSMs, whatever the mixture is. However, based on $\Phi$(B) there is a clear distinction between high- and 
low-metallicity models.  The AGSS09 and AGSS09+Ne SSMs are in excellent agreement with its 
observed value, while our high-metallicity models (GN93 or GS98) are not, away by $\sim6\sigma$. The Caffau11 
SSM of intermediate metallicity remains in marginal agreement.

We compare the GN93 SSM to the two low-metallicity ones in Fig.~\ref{Fig_fluxB8_mixtures}, where 
are shown equilbrium abundances of $^8$B (the abundance created and destroyed in an equal amount at each unit of time) 
and the cumulative neutrino flux as a function of stellar radius of the $^8$B 
disintegration,
$\phi$($^8\textrm{B}$). The figure reveals the differences between the low- and high-metallicity models above all 
arise 
from the difference in the central temperatures. With a largest T$_\textrm{c}$, the GN93 model possesses more layers 
where reactions progenitor to 
the $^8$B nuclides occur. Moreover, the hotter temperatures in these layers are also a factor favouring the progenitor 
reactions.

\subsubsection{Comparison to neutrino observations: the Borexino set}

The comparison with Borexino is shown in the right panel of Fig.~\ref{Fig_flux_mixtures} and reveals a rather different 
picture. The high-metallicity SSMs better match this set. A 
first obvious reason is the increase by Borexino of $\Phi$(B) by $\sim 10\%$ in comparison to B16. Given our 
discussion in Sect.~\ref{subsection-b16-mixtures}, the Borexino set will hence 
naturally favours SSMs with the largest T$_c$, that is the high-metallicity ones. This is indeed confirmed by the 
 values of $\chi^2_{\textrm{neutrino}}$ in the GN93 and GS98 cases. Due to the differences between the sets 
of neutrino fluxes used for the comparison, the chemical mixtures in solar models that lead to a better match to 
these observations vary. The Borexino set predicts $\Phi$(pp) and $\Phi$(Be) respectively larger by ~2\% and ~4\% 
than in the B16 analysis. However, B16 also provided estimates of these two fluxes without taking into account of the 
L$_{\odot}$ constraint, like in the Borexino approach. In that case, the two fluxes given in B16 are then respectively 
larger by ~2\% and lower by ~3\% than the Borexino ones. The inclusion of the L$_{\odot}$ constraint 
may explain the difference in the $\Phi$(pp) between the two studies, but the differences in $\Phi$(Be) and $\Phi$(B) 
(see 
also comment in Sect.~\ref{section-obs}) remain of unclear origin. The B16 study relies on a 
meta-analysis of neutrino observation data spanning on decades, and not benefiting of the latest campaigns carried out 
by the Borexino experiment posteriorly. Owing not only to the consideration of different observational datasets, 
the details of the statistical and 
physical approaches (including the neutrino oscillation parameters) used to derive the absolute neutrino fluxes are 
intricated, and could be well a source of the differences. Investigating these details are much out of the scope of the 
present work. As a perspective, we encourage for an effort in reproducing a meta-analysis like in B16, now 
including the last experimental results from the Borexino collaboration, and see whether they tend to increase the  
derived values, in particular of $\Phi$(Be) and $\Phi$(B).

The great interest for Borexino results is its successful effort for measuring with precision the neutrinos 
processed by the CNO cycle. As shown in Fig.~\ref{Fig_flux_mixtures} and despite the large errors on $\Phi$(CNO), there 
is a 
clear distinction between the SSMs of high metallicity, well within the 1$\sigma$ error bar, and the low-metallicity 
AGSS09 and AGGS09+Ne outside of it. The changes in the flux between these SSMs are linked to
abundances variations, as confirmed in Fig.~\ref{Fig_fluxCN_mixtures}, where are shown the abundances of $^{12}$C and 
$^{14}$N implied in the $\textrm{\isotope{C}{12}}(\textrm{p},\gamma)\textrm{\isotope{N}{13}}$ and
$\textrm{\isotope{N}{14}}(\textrm{p},\gamma)\textrm {\isotope{O}{15}}$ reactions. The reactions are progenitor to the 
nuclides whose 
disintegrations produce neutrinos of the CN cycle (Eqs~(\ref{eqn4}-\ref{eqn5})). The first reaction is one of the two 
fastest implied in the CNO cycle and acts as a catalyst, while the second is the slowest. In the figure, the 
cycle is at equilibrium for $\log T\gtrsim 7.06$, where most of the nuclides involved in the cycle are in 
the form of $^{14}$N, as expected from the bottleneck role of the reaction. It controls the processes of 
the CN cycle, and the production rates of neutrino  associated with this sub-cycle are therefore the same. 

In region of $\log T$ between $\sim$7.06 and $\sim$7, out-of-equilibrium reactions continue at different rates 
depending on their sensitivity to the temperature \citep[see also explanation in][]{haxton13}.  There, 
$\phi_{\textrm{\isotope{O}{15}}}$ no longer evolves, indicating the 
$\textrm{\isotope{N}{14}}(\textrm{p},\gamma)\textrm {\isotope{O}{15}}$ is not efficiently acting. The  $\Phi$(N) 
and $\Phi$(O) would be the same if $\textrm{\isotope{C}{12}}(\textrm{p},\gamma)\textrm{\isotope{N}{13}}$ was not 
burning fresh $^{12}$C from the external layers in these regions (see the gradient in the abundance of $^{12}$C). Yet, 
all the differences between the two fluxes come from these out-of-equilibrium layers, where we observe a second 
increase in $\phi_{\textrm{\isotope{N}{13}}}$. The differences in the total neutrino fluxes when varying 
the chemical mixtures clearly appears as a consequence of difference in the envelope\footnote{i.e. non 
nuclear-processed 
material} abundances of N, and more importantly of C. The GN93 SSM yields larger CNO neutrino fluxes in line with 
both its higher abundances of C and higher metallicity.

Although the trend is clear, the low-metallicity SSMs are not evidently disqualified since they remain within 
2$\sigma$ to the Borexino measure of 
$\Phi$(CNO). But it confirms the potential of this flux to test the abundances \citep[see also][]{gough19}, in 
particular if its precision could be improved in the future.

\subsubsection{Comparison to the frequency ratios }   
   
The fit to frequency ratios by the SSMs of various composition in Fig.~\ref{Fig_ratios_mixtures} illustrates with no 
discussion that only the models of high metallicity are in rather good agreement, whereas those of low metallicity are 
disqualified.  The values of the merit functions including the seismic contribution in
Table~\ref{table-mixtures} accordingly show clear decrease for the GN93 and GS98 models. These 
results confirm those initially brought out by \citet{basu07r}. \citet{chaplin07} explored in more detail the 
sensitivity of the ratios, and have shown they are particularly sensitive on the 
mean molecular weight of the core layers ($0-0.2$~$r/$R$_{\odot}$), through the dependence of ratios on the 
derivative of the sound speed, $dc/dr$ (with $c=\sqrt{\Gamma_1 P / \rho}$). We find the same origin 
to the differences in behaviour between models GN93/GS98 and Caffau11/AGSS09/AGSS09+Ne, for which marked variations in 
$dc/dr$ at the core appear between the two groups of models. The changes are correlated with differences in 
$\rho$, and so confirm the dependence to $\mu$ of this indicator in the most central layers.

   \subsection{Testing the stellar opacities}
\label{section-opacity}

\begin{table*}

\caption{Same as Table~\ref{table-mixtures} but for the SSMs calibrated with different opacity tables.}            
\centering          
\label{table-opacities}
\begin{tabular}{l l l l l }     %  columns 
\hline\hline       

            \noalign{\smallskip}

    Solar calibration & OPAL & OP & OPAS & OPLIB \\
%                 \noalign{\smallskip}
%                 & (reference calibration) & & & & \\ 
    \hline
                \noalign{\smallskip}

    X$_{\textrm{c}}$ (X$_{\textrm{0}}$) & 0.356 (0.719) & 0.358 (0.720) & 0.359 (0.723) & 0.364 (0.726) \\
                                            
                \noalign{\smallskip}        
    
    Z$_{\textrm{c}}$ (Z$_{\textrm{0}}$)  & 0.0162 (0.0151) & 0.0163 (0.0151) & 0.0164 (0.0152) & 0.0163 (0.0152) \\
    
    \noalign{\smallskip}

    T$_\textrm{c}$  [$\times 10^6$K] & 15.54 & 15.52 & 15.54 & 15.37 \\
    \hline
                \noalign{\smallskip} 
         
   $\Phi$(pp) [$\times 10^{10}$ /cm$^2$ /s] &      5.995  &  6.007  &  6.012  &  6.034   \\
               \noalign{\smallskip}                           
   $\Phi$(Be) [$\times 10^{9}$ /cm$^2$ /s]  &    4.710  &  4.681  &  4.647  &  4.354   \\
               \noalign{\smallskip}                           
   $\Phi$(B) [$\times 10^{6}$ /cm$^2$ /s]   &    5.015  &  4.935  &  4.972  &  4.090   \\
               \noalign{\smallskip}                           
   $\Phi$(N) [$\times 10^{8}$ /cm$^2$ /s]  &     2.273  &  2.252  &  2.262  &  2.027  \\
               \noalign{\smallskip}                           
   $\Phi$(O) [$\times 10^{8}$/cm$^2$ /s]   &     1.695  &  1.675  &  1.687  &  1.444  \\
               \noalign{\smallskip}                           
   $\Phi$(F) [$\times 10^{6}$/cm$^2$ /s]    &    3.619  &  3.570  &  3.601  &  3.034   \\
       \hline

                   \noalign{\smallskip}  
   P$_0$ [s] &  2178 & 2192 &  2183 & 2147   \\
    \hline 
                \noalign{\smallskip}
                                               
       $\chi^2_{\textrm{neutrino-B16}}$  &  0.033 &   0.076  & 0.066  & 1.408  \\  
       
                      \noalign{\smallskip}  
              $\chi^2_{\textrm{neutrino-Borexino}}$    & 0.118  &  0.141 &  0.153   & 0.5532 \\ 
      
                      \noalign{\smallskip}                           
     $\chi^2_{\textrm{B16+seismo}} $        & 73.024 &  95.356 & 157.843 & 23.082  \\

                        \noalign{\smallskip}                           
     $\chi^2_{\textrm{Borexino+seismo}} $        & 71.081 & 92.777 &  153.547 & 21.625  \\

    \noalign{\smallskip}
    \hline       

   \end{tabular}      
   \end{table*}

   The models in this section are calibrated on the same basis, the AGSS09 mixture and 
SF-II reaction rates, only varying the reference opacity tables; on one hand the most common for 
stellar evolution: OPAL, OP, OPLIB, on the other hand, the more specific OPAS, which is tailored for solar 
conditions.

      \begin{figure*}
\includegraphics[width=9cm]{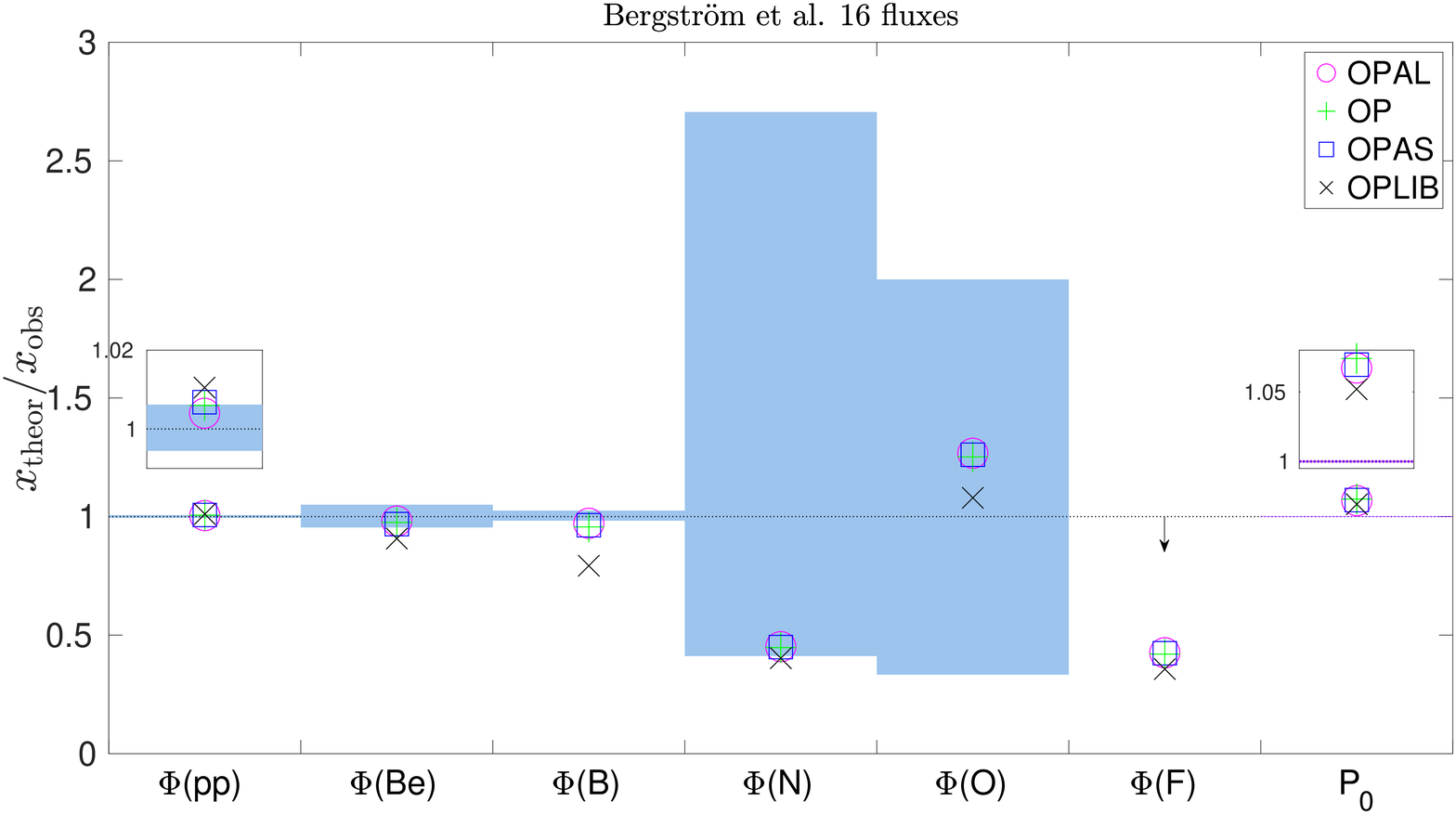}\hspace{1cm}
\includegraphics[width=9.1cm]{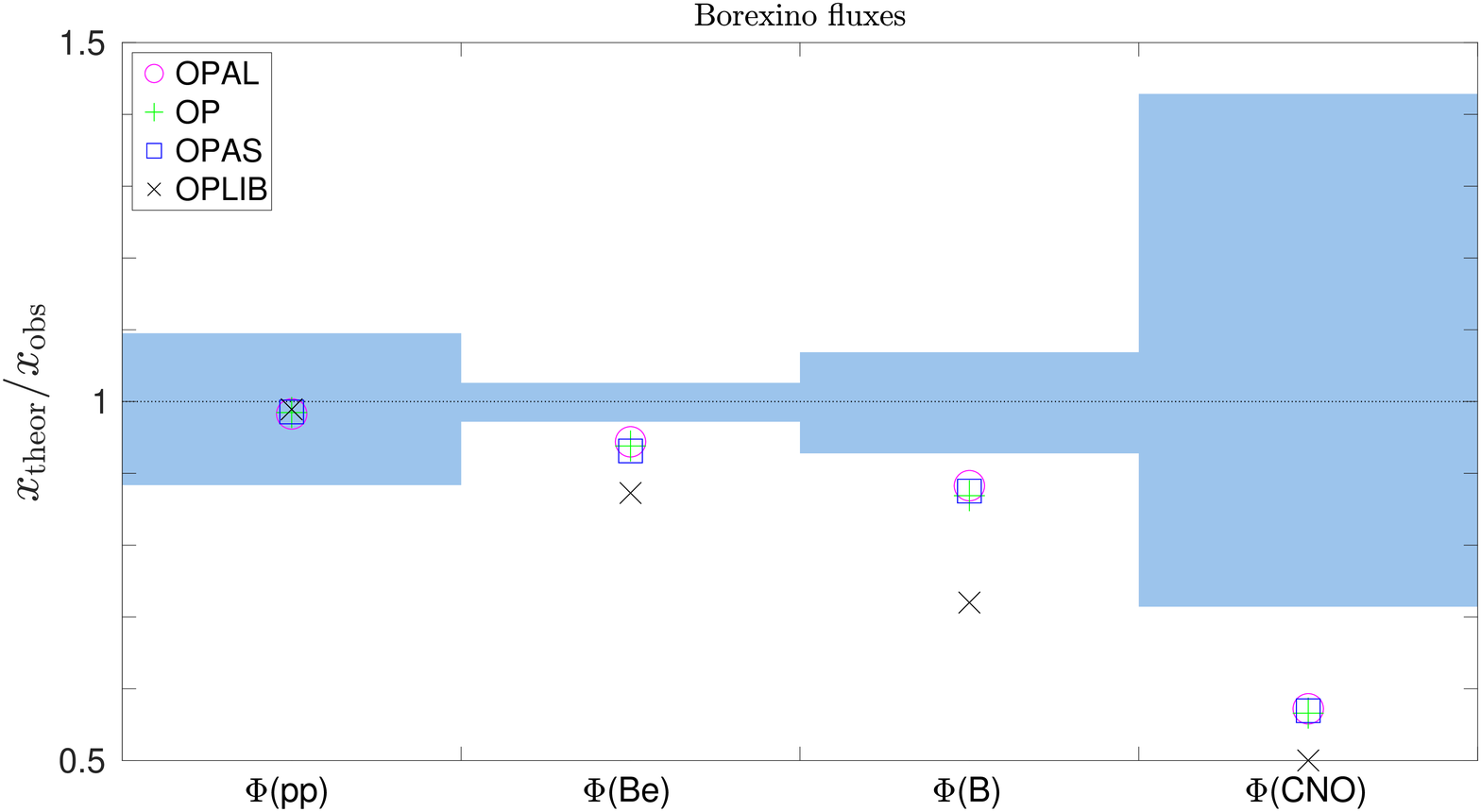}
\\
\ 
\\
\includegraphics[width=9.1cm]{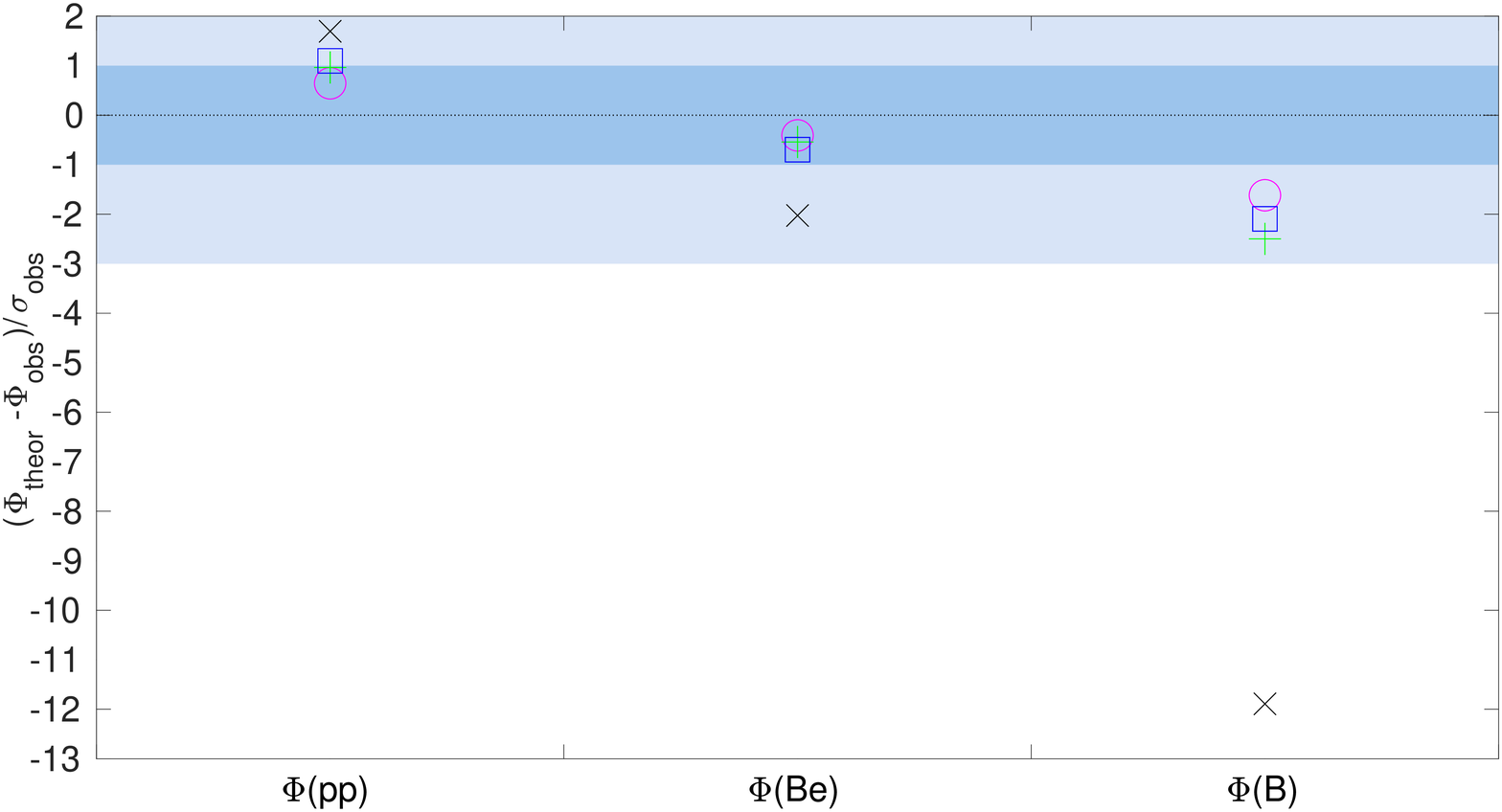}
      \caption{Same as in Fig.~\ref{Fig_flux_mixtures} but for SSMs computed with different opacity datasets.}
         \label{Fig_flux_opacities}
   \end{figure*}

In the upper left panel of Fig.~\ref{Fig_flux_opacities}, varying the opacity does not change the conclusion drawn 
in the 
previous section; the period spacings of the SSMs also discard the value reported by Fo17, by
more than 
100$\sigma$. Interestingly, the OPLIB SSM stands out because its P$_0$ is significantly reduced by 
about 40~s compared to the three other SSMs. 

\subsubsection{Comparison to neutrino fluxes and P$_0$}   

            \begin{figure*}
\includegraphics[width=19.5cm]{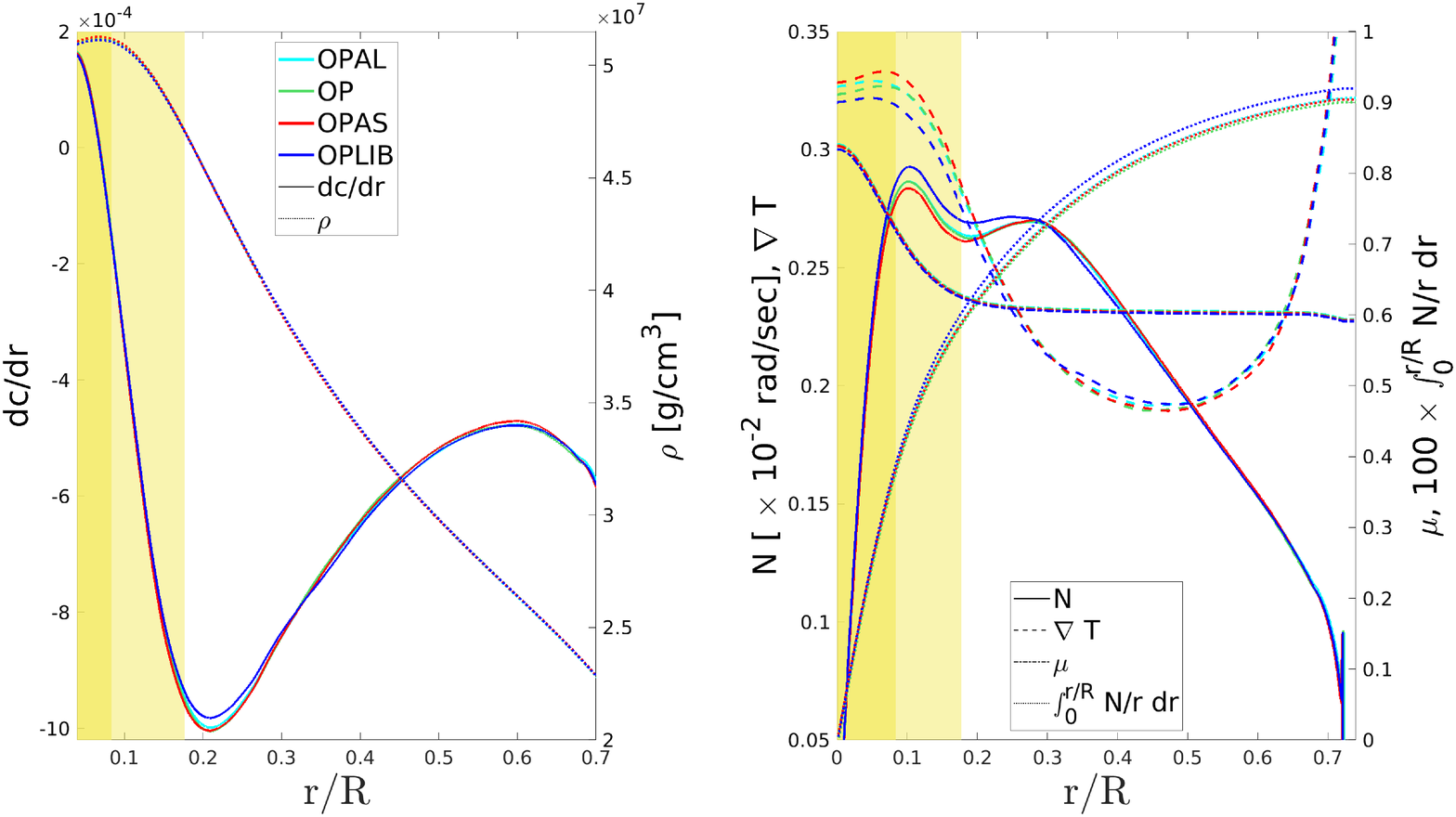}
      \caption{Comparison of acoustic variables in SSMs computed with different opacity tables, as reported in the 
legend of the left panel. The left panel shows the sound speed derivative and density as a function of the radius. 
The right panel depicts N, $\nabla T$, $\mu$, and the integral of $N/r$ along the radius. The meaning of the different 
curves in the panels are given in their respective legends. The areas shaded in yellow and light yellow respectively 
indicate the regions (starting from $r/R=0$) in which $\sim$95\% of $\Phi$(B) and $\Phi$(N) are emitted.}
         \label{Fig_acoustic_opacities}
   \end{figure*}

In Fig.~\ref{Fig_flux_opacities}, the OPLIB model fails at reproducing the fluxes whichever the 
observational set is considered. The discrepancy is maximum for $\Phi$(B), 
which is 12$\sigma$ lower than the B16 value. The departure is evidently worst with the Borexino value of this same 
flux. The 
comparison with $\Phi$(CNO) also marks a clear distinction between the OPLIB SSM and the three 
other ones. The former is half the value measured by Borexino. 

The deteriorating effect on fluxes by OPLIB were already anticipated by mean of a differential approach in 
\citet{young18}. We confirm it by a direct SSM 
calibration. The OPLIB decrease in the opacity of the solar radiative layers leads to a significant drop of 
the central temperature. To compensate and maintain the solar luminosity, the nuclear energy production by pp chain is 
increased, partly through an increase of X$_{\textrm{c}}$ (see Table~\ref{table-opacities}). According to the more 
precise data of B16, the OPLIB model overestimates production by the main pp chain. It calls for further investigation 
on the origin of the opacity decrease in OPLIB data for conditions corresponding to the solar radiative regions.

\begin{figure*}
\centering
\includegraphics[width=14cm]{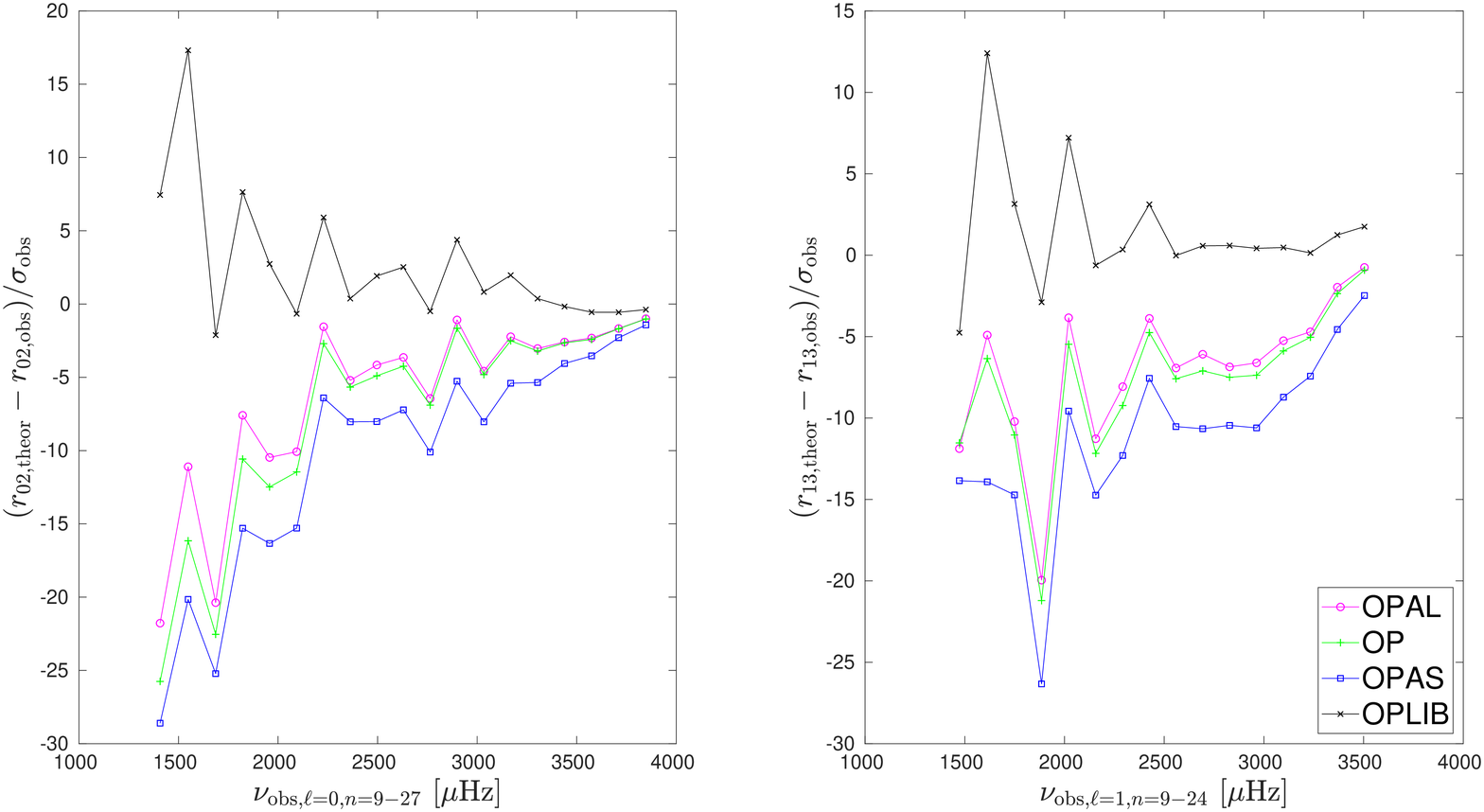}
      \caption{Same as in Fig.~\ref{Fig_ratios_mixtures} but for different adopted opacity references, as indicated 
in the legend.}
         \label{Fig_ratios_opacities}
   \end{figure*}

The effects of OP and OPAS on  X$_{\textrm{c}}$, Z$_{\textrm{c}}$ and T$_{\textrm{c}}$ in comparison to OPAL are small 
enough that they are undistinguishable from the comparisons to neutrino fluxes. As we work with the AGSS09 composition, 
we reach the same conclusion for the three SSMs, a fair agreement with the B16 data, but close to discrepancy with the 
Borexino set.  We nevertheless observe variation of 10-20~s in P$_0$ between the solar calibrations made with these 
three 
opacity datasets. With the seemingly attainable precision on the period spacing by Fo17 method, 
this indicator would be in principle useful to probe the accuracy of opacity in solar models. 

The reason for this sensitivity is easy to understand from the right panel of Fig.~\ref{Fig_acoustic_opacities}, 
where are depicted $N$, $\nabla T$, and 
the integral of $N/R$ as a function of $r$ (determining the value of P$_0$). The regions responsible for the 
distinct values of P$_0$ between the OPAL, OP and OPAS SSMs are located in $r/R<0.2$. There appear differences in 
$N$ due to the changes induced on $\nabla T$ depending on the opacity. The period spacing of the OPLIB model presents a 
larger variation because $\nabla T$ is affected by a larger amplitude and on a much larger extent of the radiative 
region, up to $r/R\sim0.65$, very close to the base of the convective zone.

\subsubsection{Frequency ratios}

The Figure~\ref{Fig_ratios_opacities} 
reveals that OP does not improve the reproduction of the ratios in comparison to OPAL, whereas OPAS worsens it. 
However, 
the OPLIB SSM significantly improves the fit. The effects of opacity are considerable since the seismic merit function 
(OPAL as reference) can be divided by a factor more than 3 by OPLIB or be twice larger with OPAS, as seen in 
Table~\ref{table-opacities}. 

The profile of  $dc/dr$ in the left panel of Fig.\ref{Fig_acoustic_opacities} explains us the reason. While it is 
similar in the OPAL, OP and OPAS models, it differs much over the whole radiative region in the OPLIB case. The 
density profiles of the four SSMs are barely affected. Therefore, the changes observed in $dc/dr$ 
arise from the differences in the temperature profile and the readjustment of the other thermodynamic quantities, in 
particular the pressure. 

The present case illustrates the need to consider as much constraints as possible to 
inspect the central structure of the Sun with solar models. Some compensatory effects can indeed lead to an apparent 
good 
agreement based on the sole seismic indicators, as with the OPLIB SSM. However, a closer inspection based on 
comparison 
with the solar neutrinos indicates a clear issue with the central physical conditions of the same model. In that 
context, the g-mode period spacing sensitivity on $\nabla T$ in the central layer emerges as an invaluable tool to 
complement these indicators. We indeed see in  Fig.\ref{Fig_acoustic_opacities} that the variables impacting its 
value vary the most (between the different SSMs) in regions located between the most central ones, which are 
better probe with help of the neutrino fluxes (see yellow regions in the figure), and the more superfical ones 
($\gtrsim0.2 r/R$), which the frequency ratios are more sensitive to.
 
\subsection{Nuclear reaction rates}
\label{compa-nuclear}

\begin{figure*}
\includegraphics[width=9cm]{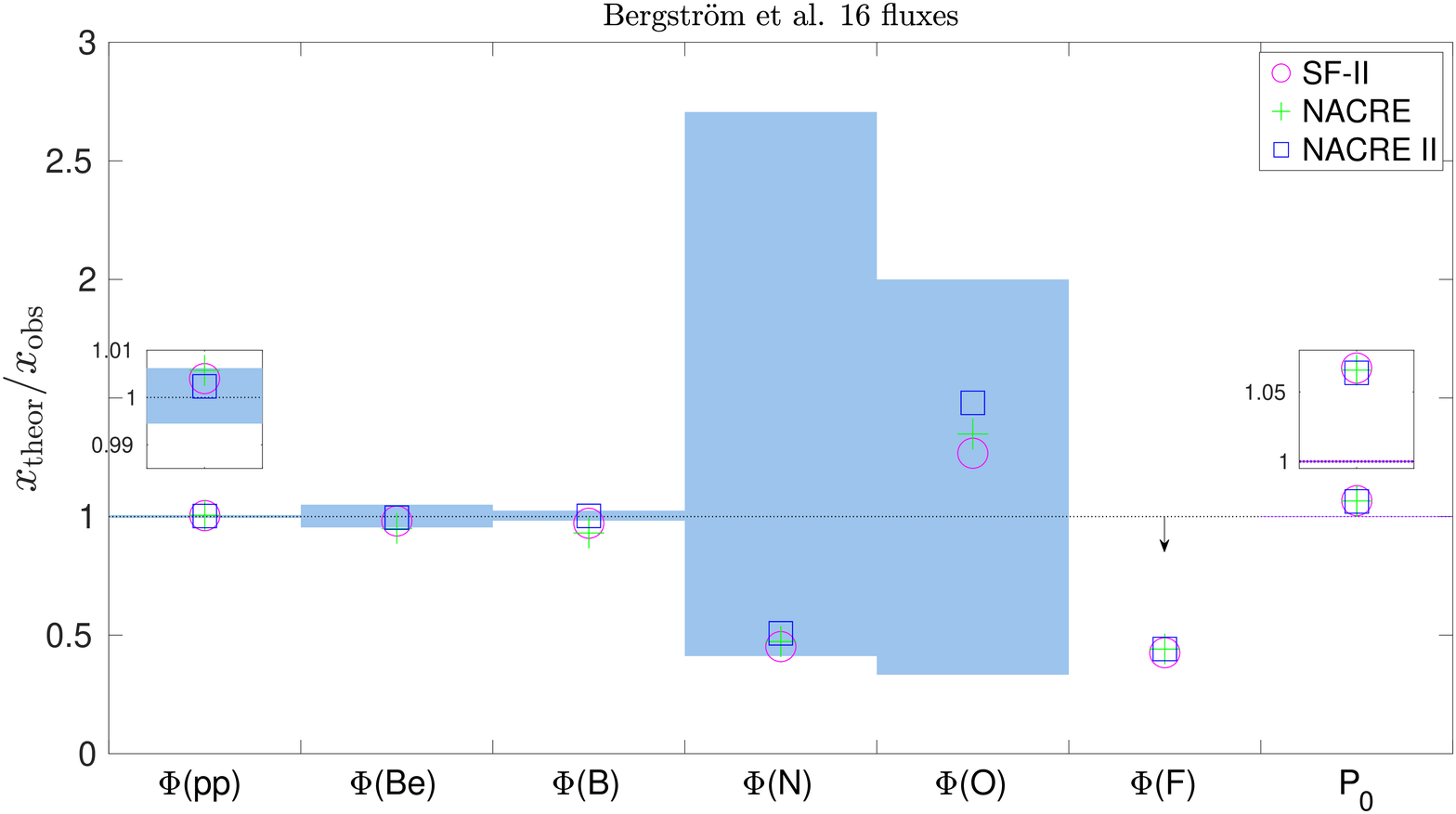}\hspace{1cm}
\includegraphics[width=9.1cm]{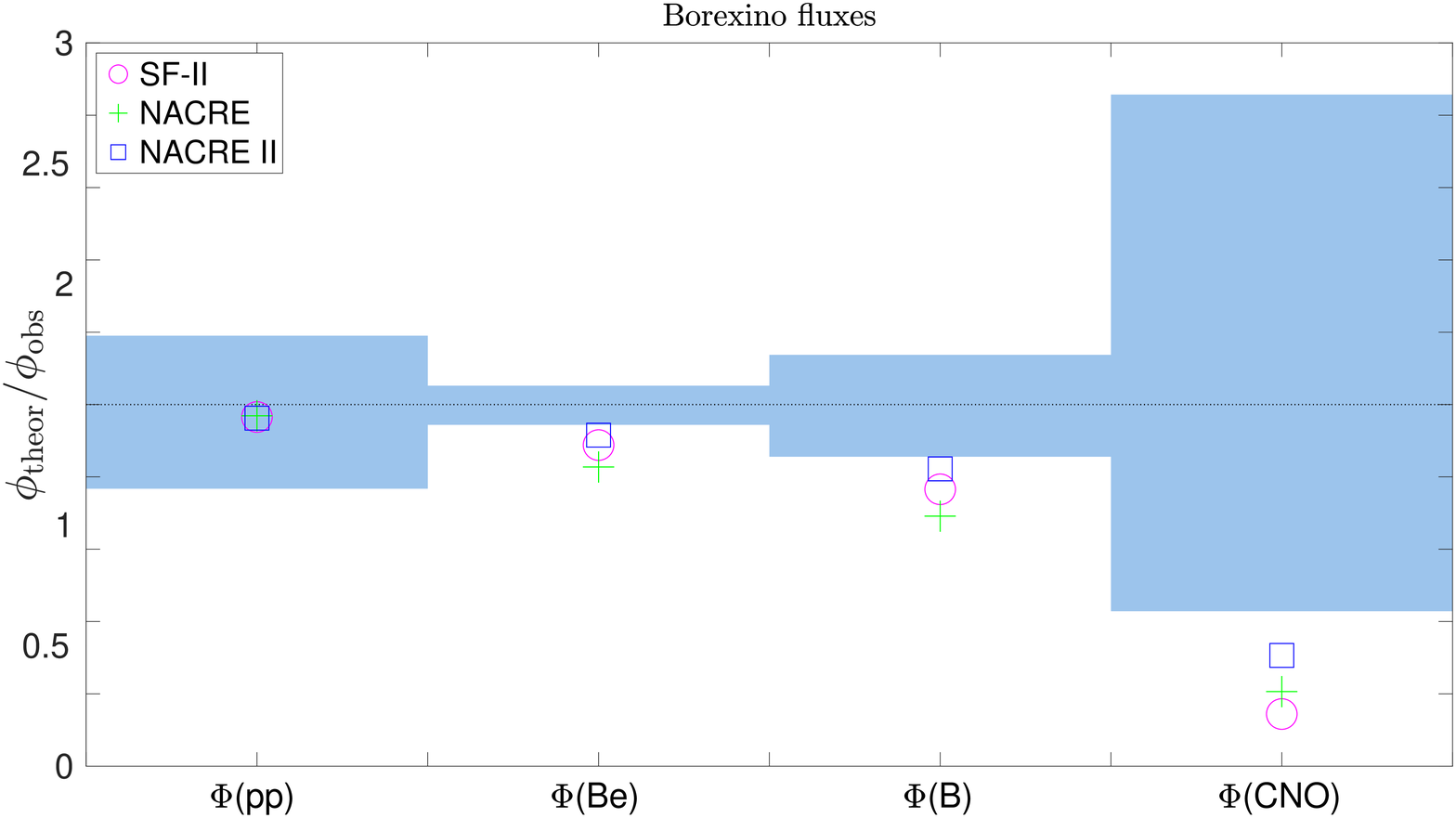}
      \caption{Same as in Fig.~\ref{Fig_flux_mixtures} (excepting the lower left panel not reproduced) but for SSMs 
computed with different sets of nuclear reaction rates. }
         \label{Fig_flux_nuclear}
   \end{figure*}

We here again adopted the AGSS09 mixture and OPAL opacities as the common ingredients, 
out of the reactions rates, of the SSMs presented in this section.
As expected, the choice of a given set of nuclear reaction rates essentially impacts the neutrino 
fluxes given by the models. First considering the three collections of rates 
detailed in
Sect.~\ref{Section_SSM}, we see in Fig.~\ref{Fig_flux_nuclear} that the NACRE SSM is the 
least accordant to the observed fluxes, either from the B16 or Borexino compilations. In particular, the 
difference with NACRE II and SF-II is due to $\Phi$(B): the underestimation of this flux is 
also the main reason for the degradation of $\chi^2_{\textrm{neutrino}}$ for this SSM in Table~\ref{table-nuclear}. 
The comparison made in \citet{xu13} between NACRE and NACRE II pinpoints at the origin of this
difference of flux: NACRE rates are lower by a few percent, and of an almost same amount over all 
temperatures,  for the $\isotope{\textrm{H}}{2}(\textrm{p},\gamma)\isotope{\textrm{He}}{3}$ and 
$\isotope{\textrm{He}}{3}(\alpha,\gamma)\isotope{\textrm{Be}}{7}$ reactions, which are both progenitors to the 
formation of $\isotope{\textrm{B}}{8}$. 

We also see in Fig.~\ref{Fig_flux_nuclear} a clear distinction between the two SSMs with NACRE II and SF-II sets 
on the Borexino CNO flux, due mostly to the difference in $\Phi$(O). From values in 
Table~\ref{table-nuclear}, the flux predicted by the NACRE II model exceeds by $\sim 16.8\%$ that of the SF-II one. A 
careful check of the models and the reaction rates reveals three sources for this large difference. The dominant term 
comes from the difference in the S-factor S(O) of the $\textrm{\isotope{N}{14}}(\textrm{p},\gamma)\textrm 
{\isotope{O}{15}}$ reaction, which differs by $\sim 8\%$ between SF-II and NACRE II. This is a consequence of distinct 
methods used for the computation of the reaction rates between the two sets, the former based on 
the R-matrix, the latter on the potential model. This leads to differences varying from $\sim$10 to 12\%, depending on 
the temperature, in the rate of this reaction between the two SSMs.  
We also find a difference of $\sim 1\%$ in the density of the radiative layers between the two models, contributing to 
increase the flux of the Nacre 2 SSM. Finally, this same SSM present a larger central temperature, which also 
contributes to an increase of $\Phi$(O).

The comparison with the Borexino $\Phi$(CNO) proves its strong potential to explore the uncertainties 
affecting the $\textrm{\isotope{N}{14}}(\textrm{p},\gamma)\textrm 
{\isotope{O}{15}}$ reaction rate; the change of this rate in Nacre 2 here
improves and shifts close to agreement at 1$\sigma$ a SSM with the AGSS09 mixture (see the right panel of 
Fig.~\ref{Fig_flux_nuclear}). Improving the precision on this flux would be helpful not only to tighten the 
central solar composition, but also to explore the rates of the CNO cycle, in particular of its bottleneck reaction.

\begin{table*}

\caption{Same as in Table~\ref{table-mixtures} but for SSMs with different reference sets of nuclear reaction rates. 
Two SSMs also include ad hoc modifications of two pp reaction rates (see main text).}         
\centering          
\label{table-nuclear}
\begin{tabular}{l l l l l l}     %  columns 
\hline\hline       
            \noalign{\smallskip}

    Solar calibration & SF-II & NACRE & NACRE II & SF-II-pp1.05-dp1.10 & SF-II-pp0.95-dp0.90 \\
%                 \noalign{\smallskip}
%                 & (reference calibration) & & & & \\ 
    \hline
                \noalign{\smallskip} 
X$_{\textrm{c}}$ (X$_{\textrm{0}}$) & 0.356 (0.719) & 0.358 (0.719) & 0.356 (0.719) &  0.357 (0.718) & 0.355 
(0.719) \\
                                            
                \noalign{\smallskip}        
    
    Z$_{\textrm{c}}$ (Z$_{\textrm{0}}$)  & 0.0162 (0.0151) & 0.0162 (0.0152) & 0.0162 (0.0152) & 0.0161 (0.0151) & 
0.0163 (0.0152) \\
    
    \noalign{\smallskip} 
                                       
    T$_\textrm{c}$  [$\times 10^6$K] & 15.54 & 15.57 & 15.57 & 15.45 & 15.65 \\
    \hline
                \noalign{\smallskip} 
         
   $\Phi$(pp) [$\times 10^{10}$ /cm$^2$ /s] &    5.995 &   6.005 &   5.985 & 6.021 &  5.965  \\
               \noalign{\smallskip}                                                                    
   $\Phi$(Be) [$\times 10^{9}$ /cm$^2$ /s]  &    4.710 &   4.559 &   4.778 & 4.474 &  4.970  \\
               \noalign{\smallskip}                                                                    
   $\Phi$(B) [$\times 10^{6}$ /cm$^2$ /s]   &    5.015 &   4.803 &   5.189 & 4.393 & 5.757   \\
               \noalign{\smallskip}                                                                    
   $\Phi$(N) [$\times 10^{8}$ /cm$^2$ /s]  &    2.273 &   2.380 &   2.554 & 2.047 &  2.548  \\
               \noalign{\smallskip}                                                                 
   $\Phi$(O) [$\times 10^{8}$/cm$^2$ /s]   &    1.695 &   1.804 &   1.980 & 1.466 &  1.974  \\
               \noalign{\smallskip}                                                                    
   $\Phi$(F) [$\times 10^{6}$/cm$^2$ /s]    &    3.619 &   3.751 &   3.754 & 3.104 &  4.248  \\
   \hline 
                   \noalign{\smallskip}  
   P$_0$ [s] & 2178 & 2175 &  2171 & 2203 & 2152   \\
    \hline 
                \noalign{\smallskip}
  $\chi^2_{\textrm{neutrino-B16}}$  &  0.033 & 0.175 &  0.003 & 0.730 & 0.418  \\   
              \noalign{\smallskip}  
              $\chi^2_{\textrm{neutrino-Borexino}}$    & 0.118  &  0.225 &  0.068   & 0.372 & 0.014  \\
             \noalign{\smallskip}                           
     $\chi^2_{\textrm{tot-B16}} $        & 73.024 &  79.608 &   77.214 & 62.498 & 90.614 \\
                   \noalign{\smallskip}                           
     $\chi^2_{\textrm{tot-Borexino}} $        & 71.081 &  77.452 &   75.134 & 60.424 & 87.705   \\
     
    \noalign{\smallskip}
    \hline       

   \end{tabular}      
   \end{table*}

\begin{figure*}
\centering
\includegraphics[width=14cm]{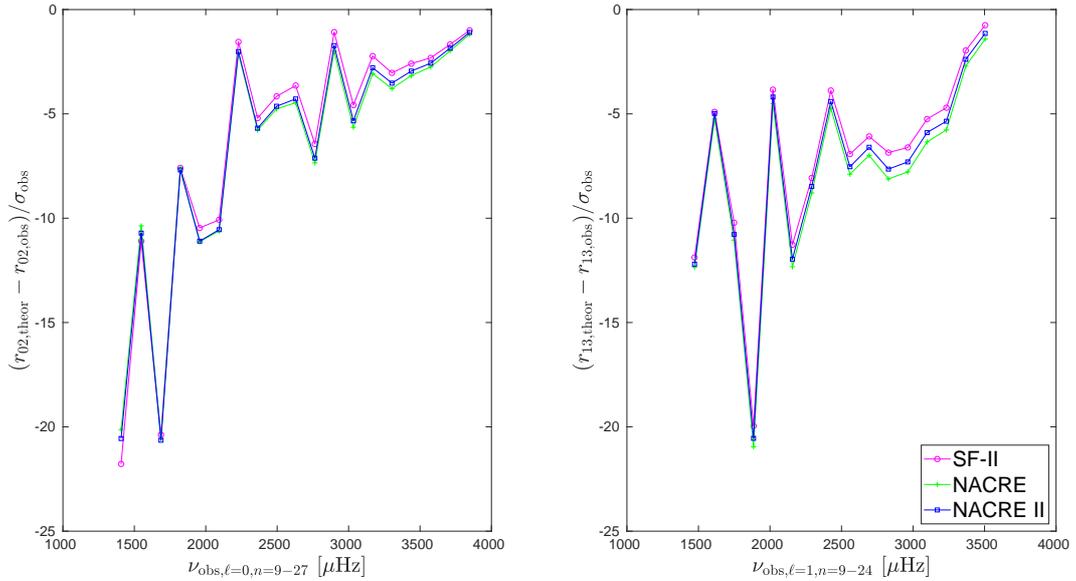}
      \caption{Same as in Fig.~\ref{Fig_ratios_mixtures} but for different reference compilations of nuclear rates, as 
indicated 
in the legend.}
         \label{Fig_ratios_nuclear}
   \end{figure*}

The helioseismic constraints are less impacted by changes in the nuclear reaction rates. The ratios are barely 
affected in Fig.~\ref{Fig_ratios_nuclear} and $\chi^2_{seismo}$ varies $\lesssim10\%$. In comparison, with a change of 
the 
composition (increase of the Ne abundance) or the opacity,  $\chi^2_{seismo}$ could be divided/increased up to a factor 
2. 
The reason was seen in the two previous sections: a change in these physical ingredients has more direct consequences 
on 
the 
central mean molecular weight or temperature, two key parameters of the seismic structure.

Similarly, P$_0$ between the three reference sets for nuclear rates are affected to a lower extent (7~s at 
most) than with other changes in the physics of the SSMs.

%Evidently, the three SSMs of this section remain 
%in complete discrepancy with the observational value of P$_0$.

\subsubsection{Parametric increase of pp reaction rates}

The impact of nuclear processes on the computation of a SSM is not limited to the selection of a reference set for the 
reaction rates. Uncertainties of various orders affect the nuclear parameters of solar models. 
We have tried to explore part of these uncertainties by computing a new series of SSM calibrations, focusing on two 
aspects. First, the uncertainties themselves on nuclear reaction rates that we tested by introducing ad hoc 
modification of the rates in a restricted set of reactions. Next, we went further on assessing the uncertainties 
on the screening effects, a major process affecting the nuclear reaction rates. This is detailed in the following 
Sect.~\ref{section-screening}.

         \begin{figure*}
\includegraphics[width=9cm]{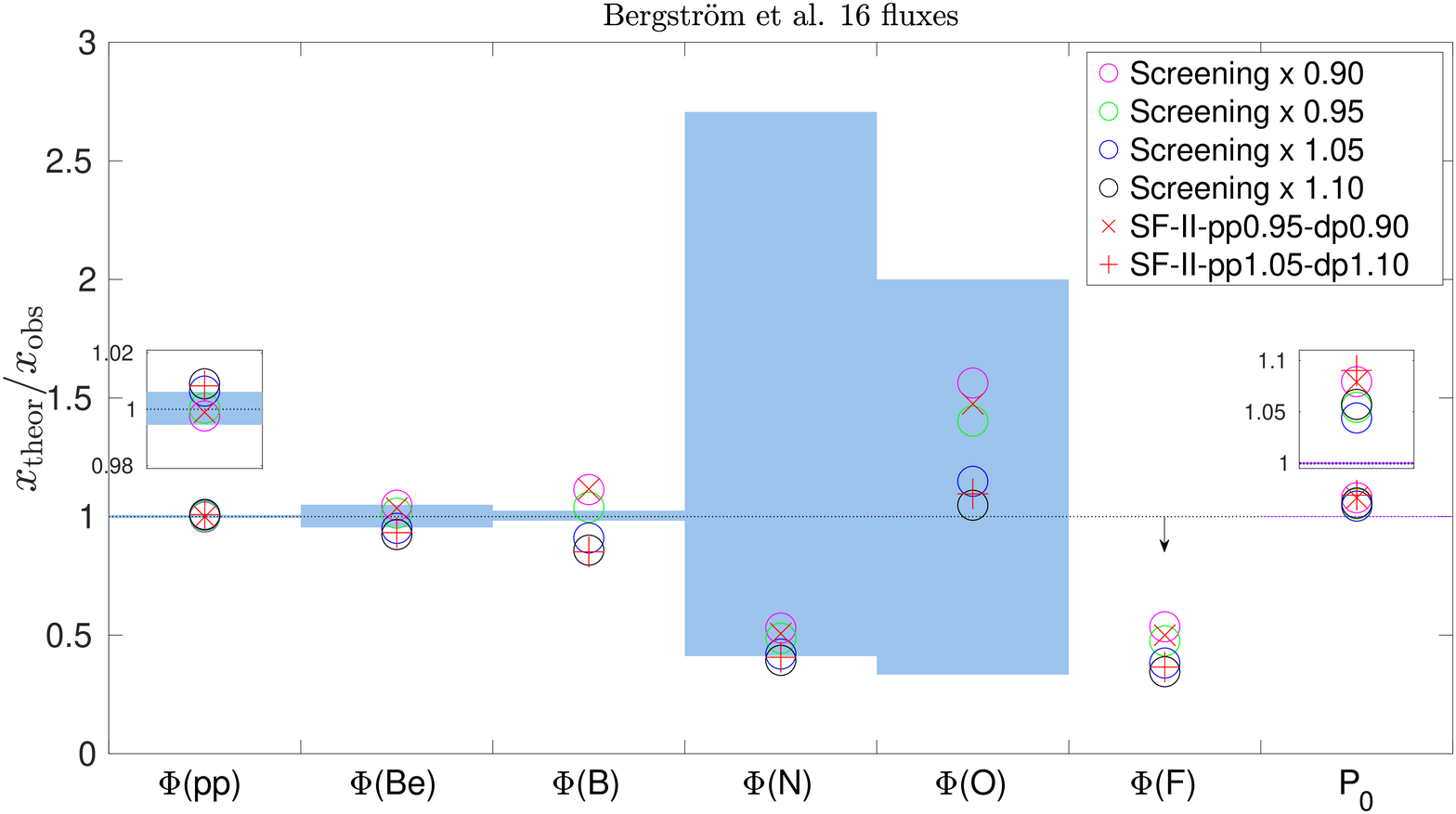}\hspace{1cm}
\includegraphics[width=9.1cm]{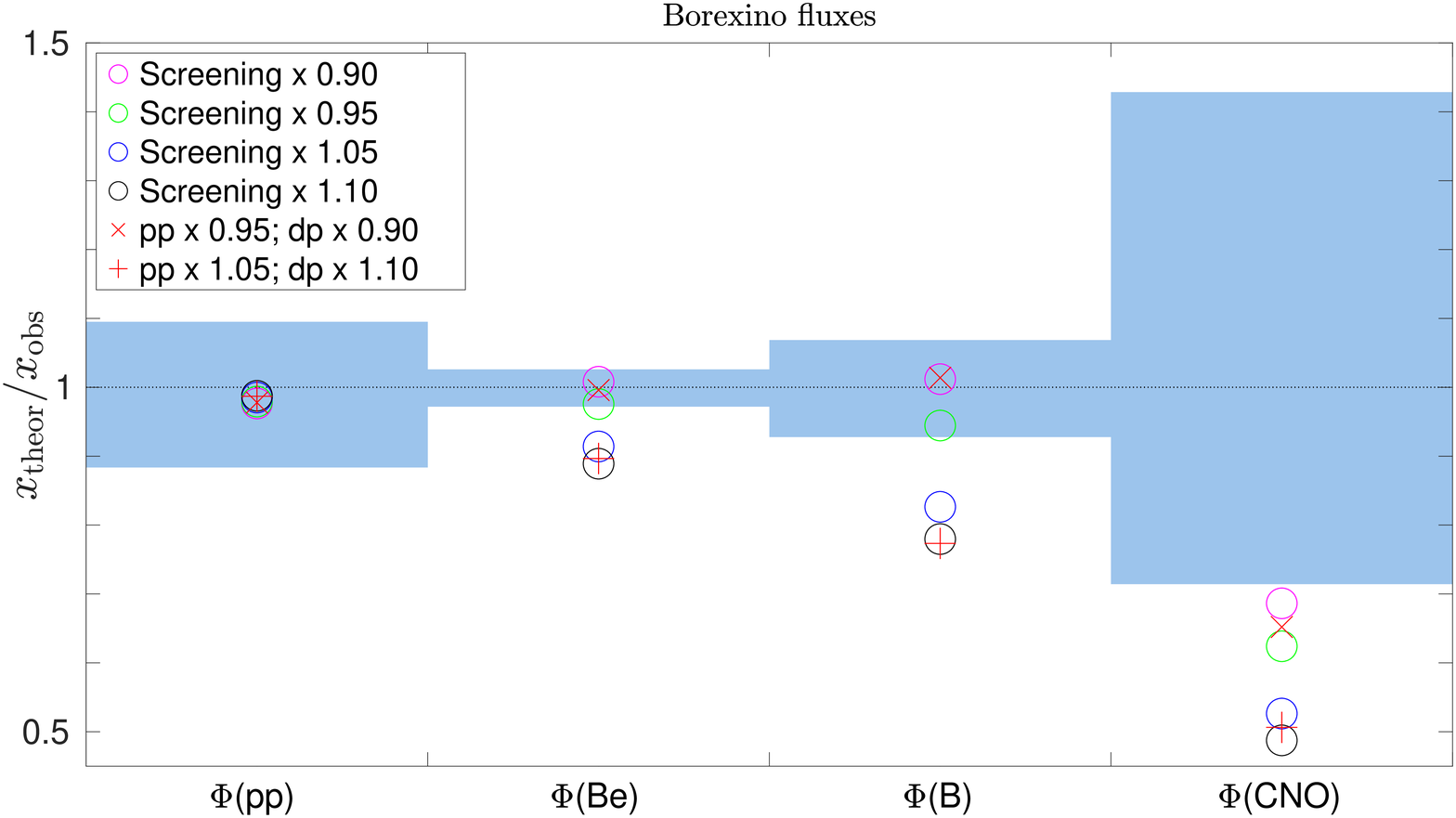}
      \caption{Same as in Fig.~\ref{Fig_flux_mixtures} (excepting the lower left panel not reproduced) but for SSMs 
computed with nuclear screening factors parametrically decreased or increased. }
         \label{Fig_flux_screenings}
   \end{figure*}
   
It is not the purpose of this work to explore the uncertainties affecting each of the reactions in the pp chains and 
CNO cycle. The interested reader will find in \citet{villante21} a recent investigation of the 
dependence of neutrino fluxes predicted by models on abundances, central temperatures and S-factors. Their work is 
based on linear pertubations of standard solar models and allows for considering dependences on an extended number of 
nuclear reactions. 
Our approach is different and cannot be as much extended since we also consider the impact of evolution on 
predictions of solar models. We hence restricted to the proton-proton (Eq.~(\ref{eqn1})) and the 
$\textrm{\isotope{H}{2}}(\textrm{p},\gamma)\textrm 
{\isotope{He}{3}}$ (hereafter d+p) reactions, for they ignite the three pp subchains. The p+p reaction rate is only 
accessible by numerical computations. \citet{adelberger11} estimate an uncertainty of about 1\% affecting its 
determination. For the second reaction, the same authors give an uncertainty around 10\%. 

Given one cannot confront experimentally the p+p computations, we exaggerated the uncertainty on it, taking into 
account 
a possible error of 5\%. By this mean, we also wanted to stress the limits of observational constraints and looked at 
whether 
they would be sensible to such extreme change. So we calibrated two additional SSMs, in the first case including an 
ad hoc increase by 5\% of the p+p simultaneously with an increase by 10\% of the d+p rate. In the second case, 
the two rates were decreased by 5 and 10\% respectively. In each case, we modify the two rates of the SF-II 
compilation, while we keep the other rates of the same compilation unaltered.  
The results are given in the last two columns of Table~\ref{table-nuclear} and in Fig.~\ref{Fig_flux_screenings} 
(red symbols). 

These modifications do not alter $\Phi$(pp) to the point of being in disagreement with the solar values of B16 or 
Borexino. However, since the calibration is done under the constraint of reproducing L$_{\odot}$, there is a 
balancing effect which leads T$_c$ to decrease when p+p and d+p rates are increased. In this SSM, as most of the 
production of energy comes from pp chains, p+p production in the models is slightly altered, because if not, the 
luminosity 
would exceed the solar one. The associated $\Phi$(pp) consequently remains within the error margins of the observed 
fluxes. However, the ratio of pp to CNO energy generation is 
much more affected and $\Phi$(CNO) presents a larger discrepancy with the Borexino data. With 
both observational set, $\chi^2_{\textrm{neutrino}}$ actually degrades significantly. This does not support increases 
of the p+p and d+p reaction rates, a result similar to \citet{ayukov17} who explored the impact of increasing the p+p 
rate. 

The SSM with the decreased rates sees its agreement with the B16 data degraded, mostly due to an increase in its 
predicted $\Phi$(B). On the contrary, its match with the Borexino set benefits of a large improvement. We 
now face a model with the AGSS09 mixture closely reproducing the dominant fluxes of the pp chains as derived with 
Borexino data. Moreover, in reason of an increase in T$_c$, the difference between the model and observed values of 
$\Phi$(CNO) reduces. In this case, testing the uncertainties of the p+p and d+p rate with the help of neutrino 
observations is ambiguous. Depending on the set of observed fluxes considered, we improve or worsen the reproduction 
of the fluxes. There again, the potential of combining neutrino with seismic constraints appears rich. Indeed, while 
the SSM with reduced rates present frequency ratio of poorer quality (but this is the case of all the models with 
AGSS09 mixture), its period spacing presents a significant drop of $\sim$25~s in comparison to models with no ad hoc 
changes. The P$_0$ of the SSM with increased rates instead increases of a similar amount. This reflects again the 
sensitivity of this indicator on the profile of temperature in the central layers of the solar models.

\subsubsection{Nuclear screening factors}
\label{section-screening}

As discussed in Sect.~\ref{Section_SSM}, the rates of nuclear reactions in the stellar models are boosted by 
the screening effect of the mean Coulomb field from the plasma. Including refined treatments of the interactions, in 
particular dynamical effects, is a tough task. Some attempts have shown that these effects could modify the screening 
factors of some reactions by a few \% \citep{shaviv07,shaviv10}. We have hence calibrated four additional SSMs, in 
which we have artificially multiplied the screening factors of all the reactions by identical factors of 0.90, 0.95, 
1.05, and 1.10. The fluxes and P$_0$ of these four SSMs are reported in Table~\ref{table-screenings}. 
\begin{table*}

\caption{Same as Table~\ref{table-mixtures} but for the SSMs calibrated with ad hoc modification in the screening 
factors.} 
           
\centering      
\label{table-screenings}
\begin{tabular}{l l l l l }     %  columns 
\hline\hline       

            \noalign{\smallskip}

    Solar calibration & Screening $\times$ 0.90 & Screening $\times$ 0.95 & Screening $\times$ 1.05 & Screening 
$\times$ 1.10 \\
%                 \noalign{\smallskip}
%                 & (reference calibration) & & & & \\ 
    \hline
                \noalign{\smallskip}
     
   X$_{\textrm{c}}$ (X$_{\textrm{0}}$) & 0.356 (0.720) & 0.356 (0.719) & 0.356 (0.718) & 0.356 (0.717) \\
                                            
                \noalign{\smallskip}        
    
    Z$_{\textrm{c}}$ (Z$_{\textrm{0}}$)  & 0.0164 (0.0153) & 0.0164 (0.0152) & 0.0161 (0.0151) & 0.0160 (0.0150) \\
    
    \noalign{\smallskip}

    T$_\textrm{c}$  [$\times 10^6$K] & 15.75 & 15.64 & 15.45 & 15.36 \\
    \hline
                \noalign{\smallskip} 
         
   $\Phi$(pp) [$\times 10^{10}$ /cm$^2$ /s] &      5.957  &  5.974  &  6.010  &  6.025   \\   
               \noalign{\smallskip}             
   $\Phi$(Be) [$\times 10^{9}$ /cm$^2$ /s]  &      5.030  & 4.869  &  4.562  &  4.437    \\
               \noalign{\smallskip}            
   $\Phi$(B) [$\times 10^{6}$ /cm$^2$ /s]   &      5.748  &  5.647  &  4.471  &  4.027   \\
               \noalign{\smallskip}            
   $\Phi$(N) [$\times 10^{8}$ /cm$^2$ /s]  &       2.668  &  2.451  &  2.115  &  1.981   \\
               \noalign{\smallskip}            
   $\Phi$(O) [$\times 10^{8}$/cm$^2$ /s]   &       2.092  &  1.876  &  1.537   & 1.402   \\
               \noalign{\smallskip}            
   $\Phi$(F) [$\times 10^{6}$/cm$^2$ /s]    &      4.550  &  4.043  &  3.251  &  2.940    \\
   \hline 
 
                          \noalign{\smallskip}  
   P$_0$ [s] & 2131  & 2158 & 2202 & 2225    \\
    \hline 
                \noalign{\smallskip}
                                               
       $\chi^2_{\textrm{neutrino-B16}}$  & 0.411  & 0.272  & 0.575 & 1.537  \\ 
                      \noalign{\smallskip}  
              $\chi^2_{\textrm{neutrino-Borexino}}$  & 0.013  &  0.035  & 0.248  & 0.397 \\  
                      \noalign{\smallskip}                    
    
     $\chi^2_{\textrm{B16+seismo}} $    &    116.364  & 91.187 & 61.806 & 57.823   \\

                        \noalign{\smallskip}                           
     $\chi^2_{\textrm{Borexino+seismo}} $        & 112.744 & 88.425 & 59.778 & 55.123 \\

    \noalign{\smallskip}
    \hline       

   \end{tabular}      
   \end{table*}

Modifying the screening factors immediately impacts the central temperature of the solar models, as seen in 
the table. A decrease (resp. increase) of the screening effect means that for a fixed set of 
temperature, density and composition, the number of nuclear reaction occurring will decrease (resp. increase). Since 
the 
total energy that must be produced by a SSM is fixed and since the amount of energy released by each reaction is 
independent, a similar total number of reactions throughout the star will be necessary to ensure the SSM radiates 1 
L$_{\odot}$. To maintain the total number of reaction, an increase (resp. decrease) of T$_c$ is required with a decrease 
(resp. increase) of screening factors. Of course some other evolutionary changes can affect the structure of the 
calibrated SSM, so that the ratio of pp- to CNO-produced energy can vary. But the dominant balance will remain ensured 
by a warming or a cooling of the core given the highly sensitive dependence of both pp and CNO nuclear energy production 
to temperature.

The $\Phi$(pp) of the four SSMs with modified screening factors cannot be discriminated by the comparison to 
observations, as the margin errors on this flux remain too large. The $\Phi$(Be) and $\Phi$(B) are more sensitive to 
T$_c$ and their values significantly differ between the SSMs. The increase of 
the screening effect reduces $\phi$(B) of the two increased SSMs in such a way that their 
disagreement with the value of B16 grows stronger, as shown in Fig.~\ref{Fig_flux_screenings}. The situation is of 
course even worse with the Borexino data, for which  $\Phi$(Be) and $\Phi$(B) predicted by the two SSMs clearly 
disagree.  
 
The tendency is reverse for the two SSMs with decreased factors. The mild decrease of 5\% restore matching between 
the fluxes predicted by an AGSS09 SSM and the B16 observations (see the left panel of Fig.~\ref{Fig_flux_screenings}). 
It needs a larger decrease of 10\% for an AGSS09 SSM to then perfectly match the Borexino $\Phi$(Be) and 
$\Phi$(B). It then also almost restore to 1$\sigma$ the agreement with $\Phi$(CNO).

The comparison to helioseismic indicators reveals an opposite situation. It is an increase of the 
screening effect that improve the fitting of the frequency ratios by AGSS09 SSMs. We observe actually the same 
behaviour as with the use of the OPLIB opacities in Sect.~\ref{section-opacity}. A decrease of the central temperature 
modifies $\nabla T$ and sees a rebalancing of other thermodynamic quantities, in particular P, so that changes in 
$dc/dr$ result in a better reproduction of the ratios. The same change of the central 
temperature considerably affects P$_0$ of the fours SSMs with modified screening factors. Its values vary by 
almost 100s between the four models, well beyond the expected observational precision.

\subsection{Equation of state and microscopic diffusion}

The microscopic processes considered in this section are the different equations of 
state currently available for solar models, and the impact 
of details in the treatment of diffusion. 

\begin{table*}

\caption{Same as in Table~\ref{table-mixtures} but for SSMs with different equations of state or microscopic diffusion 
formalisms.}            
\centering
\label{table-micro}
\begin{tabular}{l l l l l}     %  columns 
\hline\hline       
            \noalign{\smallskip}
                            
    Solar calibration & CEFF & OPAL05 & SAHA-S & Paquette coll. int.  \\
%                 \noalign{\smallskip}
%                 & (reference calibration) & & & & \\ 
    \hline          
                \noalign{\smallskip} 
X$_{\textrm{c}}$ (X$_{\textrm{0}}$) & 0.356 (0.717) & 0.355 (0.717) & 0.355 (0.717)  & 0.361 
(0.721) \\
                                            
                \noalign{\smallskip}        
    
    Z$_{\textrm{c}}$ (Z$_{\textrm{0}}$)  & 0.0162 (0.0152) & 0.0162 (0.0151) & 0.0162 (0.0151) & 
0.0158 (0.0148) \\
    
    \noalign{\smallskip} 
                                       
    T$_\textrm{c}$  [$\times 10^6$K] & 15.54 & 15.55 & 15.55 &  15.50 \\
    \hline
                \noalign{\smallskip} 

   $\Phi$(pp) [$\times 10^{10}$ /cm$^2$ /s] &   5.991 &   5.991 &   5.991 &     6.006  \\
               \noalign{\smallskip}                         
   $\Phi$(Be) [$\times 10^{9}$ /cm$^2$ /s]  &   4.723  &  4.740 &   4.737 &     4.606  \\
               \noalign{\smallskip}                         
   $\Phi$(B) [$\times 10^{6}$ /cm$^2$ /s]   &   4.998  &  5.065 &   5.049 &    4.783  \\
               \noalign{\smallskip}                         
   $\Phi$(N) [$\times 10^{8}$ /cm$^2$ /s]  &    2.277  &  2.287 &   2.286 &      2.144 \\
               \noalign{\smallskip}                      
   $\Phi$(O) [$\times 10^{8}$/cm$^2$ /s]   &    1.695  &  1.708 &   1.706 &      1.580 \\
               \noalign{\smallskip}                         
   $\Phi$(F) [$\times 10^{6}$/cm$^2$ /s]    &   3.617  &  3.649 &   3.644 &      3.360  \\
   \hline 
                   \noalign{\smallskip}  
   P$_0$ [s] & 2182 & 2174 &  2177 &  2184   \\
    \hline 
                \noalign{\smallskip}
  $\chi^2_{\textrm{neutrino-B16}}$ & 0.036  &  0.016 &   0.020  &  0.187  \\   
              \noalign{\smallskip}  
              $\chi^2_{\textrm{neutrino-Borexino}}$   & 0.114  &  0.100  &  0.103    & 0.205  \\
             \noalign{\smallskip}                           
     $\chi^2_{\textrm{tot-B16}} $      &  61.804 &  80.249 &  80.253 & 102.970 \\
                   \noalign{\smallskip}                           
     $\chi^2_{\textrm{tot-Borexino}} $      &  60.166  & 78.104  & 78.107 &  100.132  \\
    \noalign{\smallskip}
    \hline       

   \end{tabular}      
   \end{table*}

As expected, the choice of the equation of state has very little influence on the physical conditions at the centre of 
the 
models. Central temperature values for SSMs (AGSS09 mixture) with the equations of state Free, CEFF, 
OPALO5 or SAHA-S are almost identical, as presented in Table~\ref{table-micro} and \ref{table-mixtures}. The 
central compositions being almost identical between these SSMs, we do not observe any 
significant differences between their neutrino fluxes. And the conclusions we had obtained at the 
Sect.~\ref{section-mixtures} for the AGSS09 and FreeEOS SSM remain valid despite changing the equation of state. This 
lack of noticeable effect in the centre is not surprising, because as we mentioned in Sect.~\ref{Section_SSM}, the 
properties of the plasma in the core layers are in good approximation those of a perfect gas. 

A slight difference appears at the level of helioseismic indicators, for which the CEFF SSM reproduces a little better 
the frequency ratios. The CEFF equation has originally been improved for the purpose of the solar models  
to better reproduce the helioseismic data, which might explain this behaviour. The same SSM also presents 
the 
largest period 
spacing among the different equations of state; its P$_0$ is approximately 5~s higher than the 
other four SSMs, although such an increase is modest compared to the effects of other physical 
ingredients on this indicator. 

In a final test, we included departures to perfect gas description of the plasma in the microscopic diffusion 
routines. Such effects were accounted for with the help of collision integrals from \citet{paquette86}. As reported in 
Table~\ref{table-micro}, this treatment of the diffusion in a SSM with AGSS09 amplifies the difference with the 
observed neutrino datasets and the frequency ratios. 

The effect of a change in the diffusion routine is of evolutionary nature. Because the solar surface abundances are 
used as constraints, introducing more or less 
efficient settling of the elements by diffusion will require to adapt the initial composition of the solar calibration. 
It is indeed the case with the Paquette integrals, for which the SSM presents a lower initial metallicity. As 
a balancing effect, T$_c$  of the model decreases to 15.50$\times10^6$K, which is in disfavour of reproducing neutrinos 
observations.

\section{Discussion: comparison with the literature}

Some ingredients of the solar models, like the chemical mixture, play a preponderant role on the 
theoretical fluxes and orient the interpretation of solar neutrino data. Another question is how dependent it 
is on the stellar evolution code itself. For answering it, we can 
compare our SSM flux values to models with the most 
equivalent physics from other works in the literature, an exercise like e.g. \citet{boothroyd03} did for a 
previous generation of solar models.

In this aim, we have first compared our results to SSMs 
computed with the GARSTEC code \citep{weiss08} and presented in \citet{serenelli09}. We focused on the two SSMs they 
made with the  GS98 and AGSS09 mixture, which we will refer to as SSM-GS98-S09 and 
SSM-AGSS09ph-S09. The properties of these models are summarised in Table~\ref{compamodels}, while those of our two SSMs 
with the corresponding mixtures were given in Table~\ref{table-mixtures}. Their nuclear network is 
based on references presented in \citet{bahcall95}, but 
includes LUNA updates, while they use OPAL opacities and equation of state. We focus on $\Phi$(Be) and $\Phi$(B) 
because they are more sensitive to details of the core structure, in particular the temperature. These fluxes 
in our GS98 SSM are larger by 1\%, while they are lower by $\sim$2 and 4\% in the AGSS09 case. 
The $\Phi$(CNO) differs more considerably, as our SSMs estimate them larger by 
10-15\%. To the contrary, the chemical compositions at the centre are very similar; the metallicities are close by less 
than 1\% and X are <1\% in the AGSS09 case and 1.5\% in the GS98 one. These differences in $\Phi$(Be) and 
$\Phi$(B) are likely due to a mix of differences in T$_{c}$ and nuclear rates. Those affecting 
$\Phi$(CNO) are  more likely related to differences in the nuclear rates. 

\begin{table*}
\caption{Central conditions and neutrino fluxes from a selection of standard solar models in the literature (see main 
text for the references). }             
\centering  
\label{compamodels}

\begin{tabular}{l c c c c c c}     %  columns 
\hline\hline       

            \noalign{\smallskip}

    Model  & X$_c$ & Z$_c$ & T$_c$ [$\times 10^6$~K]& $\Phi$(Be) [$\times 10^{9}$ /cm$^2$ /s] & $\Phi$(B) [$\times 
10^{6}$ /cm$^2$ /s] & $\Phi$(CNO) [$\times 10^{8}$ /cm$^2$ /s]  \\ 
                 \hline
            \noalign{\smallskip} 
    SSM-GS98-S09  & 0.347 & 0.0201 & -- & 5.08 & 5.88 & 4.97\\
                \noalign{\smallskip} 
    SSM-AGSS09ph-S09 & 0.362 & 0.0160 & -- &  4.64 & 4.85 & 3.57 \\
                \noalign{\smallskip} 
    SSM-GS98-V17 & 0.347 & 0.0200 & -- & 4.93 & 5.46 & 4.88\\
                \noalign{\smallskip} 
    SSM-GS98-Zh19 & 0.349 & 0.0196 & 15.617 & 4.91 & 5.35 & 5.05 \\
                \noalign{\smallskip} 
    SSM-AGSS09-Zh19 & 0.359 & 0.0158 & 15.517 & 4.63 & 4.74 & 3.80\\
                \noalign{\smallskip}    

            \hline
         
   \end{tabular}      
   \end{table*}

In \citet{vinyoles17}, an updated GARSTEC GS98 SSM is presented, which we refer to SSM-GS98-V17 (see 
Table~\ref{compamodels}). The 
SSM-GS98-V17 shares same reference for the 
nuclear reaction rates than us, SF-II, though it is built with OP instead of OPAL in our case. The $\Phi$(Be) and 
$\Phi$(B) of our GS98 SSM are now larger by 3.7\% and 8.4\%.
The composition presents similar differences than in the above comparison, with X$_{c}$ and Z$_{c}$ respectively 1.4\% 
and 1\% higher in our model. 
The update in \citet{vinyoles17} has actually profoundly reduced the 
value they found previously for $\Phi$(B). The reason for this effect is not clear since the main element of the update 
was the revision of the nuclear reaction rates, which are now the same as the GS98 SSM we calibrated. Since  this flux 
is extremely sensitive to the core temperature, difference in T$_c$ appears as a good candidate to explain the 
important discrepancy in $\Phi$(B). 
However, referring to Sect.~\ref{section-opacity}, the effect on T$_c$ of a change of opacity from OP to OPAL 
is not sufficient to explain such a difference in $\Phi$(B). It suggests we are facing a larger difference in T$_c$ 
from 
another origin between our model 
and the SSM-GS98-V17 one. Only more detailed comparisons of 
the structures of each model could help understand the origin of this discrepancy.  

\citet{zhang19} computed with the YNEV code \citep{zhang15} SSMs with GS98 and AGSS09 mixtures (respectively the 
SSM-GS98-Zh19 and SSM-AGSS09-Zh19 in Table \ref{compamodels}), including OPAL 
opacities, and 
nuclear rates from the SF-II project, a set of physics which allow more direct comparison with our work. For instance, 
we find values of $\Phi$(B) larger by 10.6\% 5.8\% and $\Phi$(CNO) larger by 9.2 and 5.3\% 
respectively in the GS98 and AGSS09 cases. Thanks to the central temperatures of their models given in 
\citet{zhang19}, we can estimate the difference in  $T_c$ with our SSMs to be 0.38\% and 0.14\% respectively in the 
GS98 and AGSS09 case. They are of the same order as the differences in T$_c$ resulting between our SSMs when we vary 
the chemical mixtures. 

The differences are not surprising since stellar codes will intrinsically differ: internal 
error of the models, differences in the numerical schemes, but also differences in 
minor physical aspects. For instance, we realised the reference value adopted for $\textrm{L}_{\odot}$ is lower 
in our models by 0.36\% than that used in \citet{vinyoles17} and \citet{zhang19}. Therefore, we have to bear 
in mind that even with 
similar physics, fluxes predicted by different stellar evolution codes can differ by the same order than the 
differences found between sets of observed neutrinos (see B16 vs Borexino) or between models with different mixtures. 
We should be cautious that it can hence radically change our interpretation of the neutrino fluxes. It calls for more 
detailed comparisons of stellar codes to identify and quantify the distinct sources that cause them to differ.

\section{Conclusions}

%confirme doutes des observateurs

The recent announcement by \citet{fossat17} of the detection of a series of solar g modes and 
their 
rotational 
splittings has promised new constraints on the central layers of the Sun \citep[see also][on exploring 
magnetic angular momentum transport processes]{eggenberger19}. The reliability of the detection is seriously questioned 
by independent attempts to recover it \citep{schunker18,scherrer19,appourchaux19}. But, the constant spacing predicted 
by 
the asymptotic theory between the periods of g modes, P$_0$, offers in principle a valuable complement to the solar 
neutrino fluxes. Depite the fragile status of the detection, the method detailed in \citet{fossat17} and 
\citet{fossat18}, shows that the determination of P$_0$ is reachable to a high degree of precision. Adopting this 
precision, we can anticipate the constraint provided by P$_0$ on the standard solar models.

We have thus compared the theoretical P$_0$ values predicted by a set of standard solar models for which we have varied 
the main ingredients of internal physics: chemical mixture, opacity, nuclear reactions, equation of state. In 
complement to early results based on seismic models of the Sun by \citet{buldgen20}, we confirm that the 
reported g-mode period spacing is incompatible with the values predicted by standard solar models. In comparison to 
that of \citet{fossat17}, P$_0$=2041$\pm$1~s, all of the models predict a value to be larger by 100 to 150~s. 
Nevertheless, with the same level of precision, we find that it is possible to
distinguish solar models calibrated with different chemical composition, opacity, and to a lesser extent and 
with some degeneracy, screening factors and nuclear reaction rates. It is the 
sensitivity of P$_0$ to the mean molecular weight in the central layers that affords to discriminate different chemical 
mixtures and compositions. For instance P$_0$ changes by $\sim$20~s between low- and high-metallicity models. The 
sensitivity of P$_0$ to the temperature gradient also enables it to distinguish changes in the opacity. Between usual 
reference opacity sets, the value of P$_0$ typically varies by $\sim 10$~s. In complement with the frequency 
ratios of the solar pressure modes, which are also sensitive to the mean molecular weight \citep{chaplin07}, the g-mode 
period would bring helioseismology to an unprecedented level of precision to constrain the deep layers of the Sun. We 
estimate that the value P$_0$ in the Sun should most likely be between 2150 to 2190~s.

We have also compared the neutrino fluxes predicted by our models to those reported in the meta-analysis by 
\citet{bergstrom16}, and the most recent results of the 
Borexino collaboration. Although the values of \citet{bergstrom16} and Borexino are in agreement, the difference in 
the flux $\Phi$(B) is significant and can lead to a different interpretation of the comparison with models. The 
low-metallicity solar models better reproduce the B16 data while high-metallicity are preferred for the Borexino 
dataset. The $\Phi$(CNO), for the first time entirely determined in Borexino, strengthens this preference. An 
improvement in the precision on this flux, at the moment of $\sim$40\%, will clearly help to refine the question of 
abundances, in particular of metals, in the solar radiative regions. When also taking into 
account the helioseismic frequency ratios, the high-metallicity models clearly continue to be in better agreement with 
solar data \citep[confirming the results based on ratios by][]{basu07r}.

The comparison to neutrino fluxes from a standard model with the recent OPLIB opacities has shown an important 
discrepancy. 
Despite the OPLIB solar model better reproduces the frequency ratios, it is clearly incompatible with observed fluxes, 
which it underestimates by a large amount. 

We looked in detail at the potential impact of a revision of nuclear screening factors. These, in the framework 
of the so-called low-interaction regime, are currently described in most stellar models without inclusion of finer 
effects, such as those related to particle movements \citep[e.g.][]{mussack11}. These effects are 
difficult to compute and incorporate in stellar models. Nevertheless, first estimates show that factors could differ up 
to a few percent \citep{shaviv04,shaviv07}. We have tested ad hoc decreases and increases of the factors. We find 
that a decrease by 5-10\% would lead the low-metallicity models to match neutrino observations, but without improving 
their fit of the frequency ratios. This is also a consequence of the regions which the different indicators are 
sensitive to; the neutrino fluxes probe the nuclear core, the period spacing its close vicinity, where only a 
superficial number ($<5 \%$) of nuclear reactions take place, and finally the frequency ratios more superficial 
radiative regions.

This last point highlights the possibility to test multiple aspects of solar physics under consideration of the full 
observational data at our disposal. For instance, a constraint such as the present solar surface lithium abundance or 
the reproduction of all the helioseismic inverted acoustic variables are met neither with high-metallicity nor with 
low-metallicity solar models \citep[e.g.][]{buldgen20}. The solar issue remains open and likely calls for improvements 
of standard and/or non-standard stellar physics, for testing of which we benefit of an advanced set of 
observational indicators. In particular,the possibility to have constraints on the solar core from gravity modes would 
help us further refine our understanding of the solar core properties. Here, we have shown that a period spacing value 
constrained to within one or two seconds would prove very selective on the properties of solar models. In that respect, 
the quest for solar gravity modes still remains of paramount importance for studying the structure and rotation of the 
solar core.

% enlever le  paragraphe et mettre de celui d'Arlette
%Finally, we made comparison with other standard solar models from \citet{serenelli09}, \citet{vinyoles17} and 
%\citet{zhang19} computed with different stellar evolution codes. Even in the case of 
%very close physics input, significant differences in the predicted neutrino fluxes can occur. 

%In regards of the sometimes significant differences between the observed neutrino sets, the comparison with 
%results of stellar models can lead to divergent conclusions. On the one hand, we need to explore and identify the 
%sources of internal error in the models and quantify them precisely. On the other hand, continuing to improve the 
%precision and accuracy in the analysis of neutrino measurements will allow us to remove any ambiguity in their 
%interpretation. 

%In particular, these analyses are dependent on the neutrino oscillation parameters. On this last point, 
%the help of the g-mode period spacing could be valuable: it could be tried to additional 
%constraint to that of the solar luminosity when analysing neutrino data.

\begin{acknowledgements}
We thank N. Grevesse for helpful comments.
S.J.A.J.S., P.E. and G.M. have received funding from the European Research
Council (ERC) under the European Union’s Horizon 2020 research and innova-
tion programme (grant agreement No 833925, project STAREX).
G.B. acknowledges fundings from the SNF AMBIZIONE
grant No. 185805 (Seismic inversions and modelling of transport processes in
stars) and support by the ISSI team ``Probing the core of the Sun and the stars'' (ID 423).
\end{acknowledgements}

\bibliographystyle{aa} % style aa.bst
\bibliography{neutrinos}

\end{document}